\documentstyle[12pt,aaspp4]{article}

\newcommand{\asca}{{\sl ASCA} }
\newcommand{\rosat}{{\sl ROSAT} }
\newcommand{\ginga}{{\sl Ginga} }
\newcommand{\cgro}{{\sl CGRO} }

\begin{document}

\title{ASCA Observations of Blazars and Multiband Analysis} 

\author{H. Kubo\altaffilmark{1}, T. Takahashi\altaffilmark{2}, 
G. Madejski\altaffilmark{3,4}, M. Tashiro\altaffilmark{5},  
F. Makino\altaffilmark{2}, S. Inoue\altaffilmark{6}, \\and F. Takahara\altaffilmark{7}}

\altaffiltext{1}{The Institute of Physical and Chemical Research (RIKEN), 
2-1 Hirosawa, Wako, Saitama, 351-0198, Japan}
\altaffiltext{2}{Institute of Space and Astronautical Science, Sagamihara, 
Kanagawa, 229-8510, Japan}
\altaffiltext{3}{Laboratory for High Energy Astrophysics, Code 662, 
NASA GSFC, MD 20771, USA}
\altaffiltext{4} {University of Maryland, College Park, MD 20742, USA}
\altaffiltext{5}{Department of Physics, University of Tokyo, Bunkyo-ku, 
Tokyo, 113-0033, Japan}
\altaffiltext{6}{Tokyo Metropolitan University, 1-1 Minami Osawa, Hachioji, 
192-0397, Japan}
\altaffiltext{7}{Department of Earth and Space Science, Graduate school 
of Science, Osaka University, 1-1 Machikaneyama, Toyonaka, Osaka, 560-0043, Japan}

\begin{abstract}
We present data for 18 blazars observed with the X--ray satellite 
{\sl ASCA}, half of which were also observed contemporaneously with 
the EGRET instrument onboard {\sl Compton Gamma-ray Observatory} 
as parts of multi-wavelength campaigns.  The observations show a 
clear difference in the spectra between three 
subclasses of blazars, namely the High-energy peaked BL Lac objects (HBLs), 
Low-energy peaked BL Lac objects (LBLs), and quasar-hosted blazars (QHBs).  
The \asca X--ray spectra of HBLs are the 
softest, with the power law energy index $\alpha \sim 1 - 2$, 
and they form the highest observable energy tail of the low energy 
(LE, synchrotron) component.  
The X--ray spectra of the QHBs are the hardest ($\alpha \sim 0.6$)
and are consistent with the lowest observable energy end of 
the high energy (HE, Compton) component.  For LBLs, the X--ray spectra are 
intermediate.
We find that the radiation process responsible for the HE peak for HBLs 
{\sl can} be explained solely by Doppler-boosted Synchrotron-Self-Compton 
(SSC) emission, with the Doppler factor $\delta$ consistent with the 
VLBI and/or $\gamma$--ray variability data.  For many QHBs, on the other 
hand, the $\gamma$--rays {\sl cannot} be solely due to the SSC mechanism 
unless $\delta$ is significantly in excess of that inferred from 
VLBI data.  We consider an alternative scenario consistent with the measured 
values of $\delta$, where the SSC component is still present in QHBs and 
it dominates in the X--ray band, but it is below the observed $\gamma$--ray 
spectrum.  With an assumption that the peak of the SSC emission is on the 
extrapolation of the X--ray spectrum, and adopting $\delta$ of 10, 
we infer the magnetic field $B$ to be 0.1 -- 1 Gauss, and 
Lorentz factors $\gamma_{b}$ of electrons radiating 
at the peak of the $\nu F(\nu)$ spectrum of $\sim 10^{3}$ for QHBs;  this 
is much lower than $\gamma_{b} \sim 10^{5}$ for HBLs, even though the 
values of $B$ are comparable in the two sub-classes.  This difference 
of $\gamma_{b}$ is most likely due to the 
large photon density expected in QHBs (e.g. from thermal components visible 
in these objects) as compared with that of HBLs;  Compton upscattering of 
these photons may well provide the observed GeV flux.  

\end{abstract}


\section{Introduction}

The overall electromagnetic spectra of blazars -- a class of active 
galactic nuclei (AGNs) that includes BL Lac objects and Optically Violently 
Variable (OVV) quasars -- are believed to be dominated by Doppler-boosted 
radiation from relativistic jets pointing closely to our line of sight 
(\cite{blandford78}; \cite{blandford79}; \cite{urry95} for a review of 
radio loud AGNs).  
The VLBI studies of these objects show compact radio cores on milli-arcsecond 
angular scale with jet-like structures which often show superluminal motion, 
with apparent speeds $v/c \sim 5 - 10$ (e.g., \cite{vermeulen}).  
Apparent variability time scale and luminosity amplification depend on various 
powers of the ``beaming'' (Doppler) factor $\delta$ (e.g., \cite{lind}), 
defined via $\delta  = \Gamma_{j}^{-1} (1 - \beta \cos \theta)^{-1}$, 
where $\Gamma_{j}$ is the bulk Lorentz factor of the emitting matter, 
$\beta = v/c$, 
and $\theta$ is the angle of motion with respect to the line of sight.  

Blazars are commonly detected as $\gamma$--ray sources. The EGRET instrument 
onboard the {\sl Compton Gamma-Ray Observatory} (\cgro) has so far detected
emission in the GeV range from $\sim$ 50 blazars (\cite{fichtel94}; 
\cite{thompson95}; \cite{mattox97}; \cite{mukherjee97});
$\gamma$--ray emission has been detected up to the TeV range from the
nearby BL Lac objects Mkn~421 and Mkn~501 with ground-based Cherenkov 
telescopes (\cite{punch92}; \cite{petry93}; \cite{quinn96}; 
\cite{bradbury97}).  
As these sources show large luminosity and compact emission regions 
in the spectral range where the opacity to pair production via 
$\gamma\gamma\rightarrow e^{+}e^{-}$ is large, 
it is generally accepted that the $\gamma$--ray emission is anisotropic 
and Doppler-boosted as well (\cite{maraschi92}; \cite{mattox93}; 
\cite{dondi95}; \cite{buckley96}), suggesting that the {\sl entire} observed 
electromagnetic emission arises in the jet.  
As these objects emit in practically every observable waveband, any 
study of the structure and physical conditions in the jets requires broad-band 
spectral observations, which, given the rapid large amplitude flux 
variability, must be conducted simultaneously.  

The overall spectra of blazars have two pronounced components: 
one peaking at low energies (LE), $10^{13}-10^{17}$ Hz 
(e.g., \cite{sambruna96}), 
and another peaking at high energies (HE), in the $\gamma$--rays 
(e.g., \cite{montigny95}). 
For the blazars that are hosted in quasars (QHBs), 
and for BL Lac objects discovered 
via radio-selection techniques (the so-called ``Low-energy peaked BL Lacs'' 
or LBLs), the LE component peaks in the infrared. 
For the the majority of BL Lac objects 
-- those found as a result of their X--ray emission -- 
it peaks in the ultraviolet or even in the soft X--rays 
(\cite{giommi95}; \cite{sambruna96}; \cite{padovani96}; \cite{fossati97a}),
and thus they are named ``High-energy peaked BL Lacs (HBLs)'' 
(\cite{padovani95}).  
The local power-law shape, the smooth connection of the entire radio - to 
- UV (and, for the HBLs, soft X--ray) continuum, as well as the relatively 
high level of polarization observed from radio to the UV, imply 
that the emission from the LE component is most likely produced via the
synchrotron process of relativistic particles radiating in magnetic field.
This view is strongly supported by spectral variability observed in 
X--rays in a number of HBLs, where the variability at lower X--ray energies 
lags behind the more energetic X--rays (\cite{tashiro92}; \cite{sembay93}; 
\cite{kohmura94}; \cite{tashiro95}; \cite{takahashi96})

The HE component, on the other hand, peaks in the $\gamma$--ray band, in 
the MeV - to - GeV range, and, in the case of a few HBLs, it extends to the
TeV range;  it is believed to be produced via 
Comptonization by the same particles that radiate the LE component.
The source of the ``seed'' photons, 
can either be the synchrotron radiation, internal to the jet -- as in
the Synchrotron-Self-Compton (SSC) models 
(\cite{rees67}; \cite{jones74}; \cite{konigl81}; 
\cite{ghisellini85}; \cite{band85}; \cite{ghisellini89}; \cite{maraschi92}; 
\cite{bloom96}; \cite{mastichiadis97}).
Alternatively, these can 
be external to the jet, as in the External Radiation Compton (ERC) models:  
either the UV accretion disk photons (\cite{dermer92}; \cite{dermer97}), 
or these UV photons reprocessed by the emission 
line clouds and/or intercloud medium (\cite{sikora94}; \cite{blandford95}), 
or else, IR radiation ambient to the host galaxy (\cite{sikora94}).
The ratio of the power in the HE to the LE components is 
systematically larger for QHBs than for BL Lac objects 
(\cite{maraschi94a}; \cite{dondi95}; \cite{sambruna97a}; \cite{ulrich97}; \cite{fossati97b}). 

If we assume that the LE component 
is due to the synchrotron radiation, its peak frequency is 
determined by the intensity of magnetic field and the distribution function 
of electron energies, while the location of the HE peak is related to 
the distribution functions of electron and target photon energies.  
The ratio of the luminosity of these components 
($L_{HE}/L_{LE}$), in the context of this synchrotron plus Compton model, 
is expected to reflect the ratio of energy densities of photon and 
magnetic fields in the jet.  

This paper reports the X--ray spectra of 18 blazars measured by \asca 
in the context of their multi-band emission.  The \asca observations and 
results are described in \S2, followed by multi-band analysis and discussion 
in \S3. Summary of this paper is presented in \S4.
Throughout this paper we use $H_0$=75 km s$^{-1}$ Mpc$^{-1}$, $q_0$=0.5.

\section{\asca observations and multiband results}

X--ray satellite \asca (\cite{asca}) observed 18 blazars (cf. Table 1), 
half of which were also observed contemporaneously with EGRET 
as parts of multiband campaigns;  such simultaneous observing campaigns 
are essential for these highly variable objects. 

We applied the standard screening criteria to the data.
The minimum cutoff rigidity is 6 GeV/c.
The minimum elevation angle is 10 and 5 degree respectively 
for the Solid-state Imaging Spectrometer (hereafter SIS; \cite{burke91}; \cite{yamashita97}) 
and the Gas Imaging Spectrometer (GIS; \cite{ohashi96}; \cite{makishima96}). 
The minimum angle between the target and the illuminated Earth's limb is 
20 degree for the SIS. We rejected the data obtained during the passages 
through the South Atlantic Anomaly. 
We extract the source 
counts from a region of a radii of 3 arcmin and 4--6 arcmin respectively. 
For the GIS we chose a radius in proportion 
to the brightness of a source as long as the ratio of signal to background 
is one. The background counts were taken from a source-free region in 
the same image as the target for a source whose source count rate is 
below 0.2 cts s$^{-1}$. 
For sources whose count rate is above 0.2 cts s$^{-1}$, there 
is no image region where contamination from the source is negligible. 
For those cases, we took background from the same detector 
region as the location of the source, but from blank sky observations.

We fitted the spectra using 
the detector response v1.1$\alpha$ for SIS and v4.0 for GIS, 
by a power law with absorption at low energies 
(cross section from \cite{morrison}). 
In most cases, the data were well-described by this model, 
where the absorption was consistent with the Galactic value, 
although for HBLs the spectra are convex and this model is too simple 
(\cite{sambruna94}; \cite{tashiro95}; \cite{takahashi96}; 
\cite{sambruna97b}; \cite{kubo97}). 
But in all cases, a power law model with Galactic absorption 
suffices in the 2 - 10 keV range.  
When there is more than one \asca observation during 
one viewing period of EGRET,
we used an index determined from the summed data 
over the period of contemporaneous observations with EGRET. 

Figure 1 shows the distribution of the 2 - 10 keV energy indices 
of the blazars obtained with {\sl ASCA}.  This figure clearly shows that the 
X--ray spectra of HBLs are the softest, with the power law energy index 
$\alpha \sim 1 - 2$, and they form the highest observable energy tail 
of the LE component.  X--ray spectra of the QHBs are the hardest 
($\alpha \sim 0.6$) and are consistent with the lowest observable 
energy end (``onset'') of the HE component.  For LBLs, the X--ray spectra are 
intermediate;  in one case, 0716+714, the \asca spectrum shows hardening 
with an increasing energy (Fig. 2), 
similar to the spectrum observed with \rosat
(\cite{cappi94}). The differences in the spectra between three subclasses 
were also observed with \rosat in the soft X--ray band 
(\cite{sambruna96}; \cite{urry96}; \cite{sambruna97a}; \cite{padovani97}; 
\cite{comastri97}). 
The time variability in \asca observations is 
reported in Kubo (1997). It is worthwhile to note that while the energy 
index for HBLs is often variable on a short time scale, 
for QHBs the index remains almost constant when the flux changes. 

We constructed the multiband spectral energy distribution 
ranging from radio to $\gamma$--ray band for 18 sources observed 
with \asca using above data, published results, the database and the data 
taken by private communication.
The $K$-corrected multiband spectra are shown in Fig. 2. 
For the $K$-correction, the flux densities are multiplied by 
$(1+z)^{\alpha-1}$, where $\alpha$ is the energy spectral index at 
frequency $\nu$, defined such that flux $F(\nu)\propto\nu^{-\alpha}$. 
We used individual spectral indices when available, otherwise average 
indices for each subclass of blazars (\cite{sambruna96}).
We plot the simultaneous data when available, and 
for non-simultaneous data we averaged the flux densities at a given frequency.
In order to determine the peak frequency and the 
luminosity of both the LE and HE components, we performed a empirical 
third-order polynomial fit to the spectra in $\nu F (\nu)$ (\cite{comastri95}).
The results are listed in Table 2. Figures 4a and 4b show that, for QHBs,  
the peak frequency of the LE component is lower, 
$L(LE)$ and the $L(HE)/L(LE)$ ratio is greater 
than that measured in HBLs. 
For LBLs, these properties are, again, intermediate.

\section{Discussion}

As we mentioned previously, the two leading models of the high energy 
emission in blazars invoke Comptonization, of internal (SSC) or external 
(ERC) seed photons.  In the following analysis, we assume that {\sl both} SSC 
and ERC processes may operate in blazars.  We then estimate the contribution 
of the SSC emission in the HE component.  In order to calculate the 
predicted luminosity due to the SSC emission, we assume a simple 
homogeneous model, in which photons are produced in a region of radius 
$R$ and with a constant magnetic field $B$.  

We considered the radiation by a single population of relativistic 
electrons, with a broken power law distribution of Lorentz factors 
$\gamma_{el}$, and a break point at $\gamma_{b}$ (similar to e.g. 
\cite{sambruna96}).  We also assume that the radiation spectrum of 
the LE component peaks at a frequency corresponding to that radiated by 
the electrons with $\gamma_{b}$.  

The peak frequency of the synchrotron component in the observer frame, 
$\nu_{sync}$, is then given as, when pitch angle is $\pi$/2: 
\begin{equation}
\nu_{sync} = 1.2 \times 10^{6} \gamma_{b}^{2} B \frac{\delta}{(1+z)} \quad {\rm Hz}
\label{eqn1}
\end{equation}

where B is in Gauss.  If the electron energy is still in the Thomson regime, 
($\gamma_{el} \times h\nu_{sync} << m_ec^2$),
the expected peak of the SSC component in the observer frame 
($\nu_{SSC}$) is 
$\nu_{SSC} = 4 \gamma_{b}^{2} \nu_{sync}/3$.
The ratio of the observed luminosity of the SSC component 
$L_{SSC}$ to the observed synchrotron luminosity 
$L_{sync}$ is: 
\begin{equation}
\frac{L_{SSC}}{L_{sync}}=\frac{u_{sync}}{u_{B}}
\label{eqn2}
\end{equation}
where the $u_{sync}=L_{sync}/(4\pi R^{2}c\delta^{4})$ is the 
rest-frame energy density of the synchrotron photons, and 
$u_{B}=B^{2}/(8\pi)$ is the magnetic field energy density.  

To check the validity of the assumption that the observed HE component is 
solely due to the SSC emission, we calculated the beaming factor ($\delta$), 
which is given from above equations: 
\begin{equation}
\delta^2 = 1.6\times 10^{12} \frac{L_{sync}}{c R^2} 
\left(\frac{L_{sync}}{L_{SSC}}\right) \frac{\nu_{SSC}^2}{\nu_{sync}^4} \frac{1}{(1+z)^2}
\label{eqn3}
\end{equation}

where $L$ is in erg~s$^{-1}$, $\nu$ in Hz, $c$ in cm s$^{-1}$, and $R$ in cm.  
We estimate $R$ from the 
shortest observed variability (doubling) time scale $\Delta t$ 
observed in any wavelength, as given in Table 2.  Assuming that 
$R \lesssim c \delta \Delta t / (1+z)$, then Eq. 3 can be rewritten as:
\begin{equation}
\delta^4 \gtrsim 1.6\times 10^{12} \frac{L_{sync}}{c^3 \Delta t^2} 
\left(\frac{L_{sync}}{L_{SSC}}\right) \frac{\nu_{SSC}^2}{\nu_{sync}^4}
\label{eqn4}
\end{equation} 

where $\Delta t$ is in s, and other quantities are as in Eq. 3.  
An application of this equation to the data in Table 2 
assuming $L_{sync}=L_{LE}$, $L_{SSC}=L_{HE}$, $\nu_{sync}=\nu_{LE}$, $\nu_{SSC}=\nu_{HE}$ 
implies that the lower limits of $\delta$ for HBLs are $\sim $3 or less,
which is consistent with the VLBI results (cf. \cite{vermeulen}), 
and the limits obtained from the arguments of the $\gamma$--ray opacity 
(cf. \cite{dondi95}).  However, for 4 QHBs, where the $\gamma$--ray 
flux severely dominates the radiative output, we derive values of 
$\delta$ that are much larger than the VLBI results (see Table 2).  This 
suggests that an additional emission mechanism -- such as the ERC process -- 
may contribute significantly in the $\gamma$--ray 
regime, dominating over the SSC flux, and the values of $\nu_{SSC}$ and 
$L_{SSC}$ are {\sl very} different than $\nu_{HE}$ and $L_{HE}$, with
the SSC component ``hidden'' well below the ERC component.  

However, the fact that the QHBs have X--ray spectra which are hard, with 
$\alpha \sim 0.6$, and which are {\sl not} located on the extrapolation of the 
synchrotron optical / UV spectra, implies that the X--rays observed 
in QHBs are due to a separate emission process than synchrotron.  
The fact that for most of QHBs the $\gamma$--ray spectra are above 
the extrapolation of X--ray spectra (Fig. 2) suggests that the 
dominant process is different for X--rays than it is for $\gamma$--rays.  
One explanation is that the SSC process dominates in the X--ray range, 
while the ERC process dominates in $\gamma$--rays (\cite{inoue}). 
With the assumption that SSC process is dominant in X--rays for QHBs,
we estimate the location of the ($\nu_{SSC}$, $L_{SSC}$) point 
in the log($\nu$) -- log($\nu F (\nu)$) space by the following method.  
We assume that it lays on or below the extrapolation of the {\sl ASCA} 
spectrum (line (a) in Fig. 3), but above the highest value of $\nu F (\nu)$ 
measured by {\sl ASCA}. Since the spectra of QHBs generally have 
$\alpha < 1$ and thus $\nu F (\nu)$ is the highest at the end of the 
\asca bandpass (10 keV $\simeq$ 2$\times$10$^{18}$Hz), this second 
limit is equivalent to 
$\nu_{SSC} L_{SSC} > 2\times 10^{18}$ Hz $L_{10 \rm keV}$ (line (b) in 
Fig. 3).  
We further constrain $L_{SSC}$ using Eq. 3;  
once we assume a given $\delta$, there is a unique relationship 
between $L_{SSC}$ and $\nu_{SSC}$ described as:  
\begin{equation}
L_{SSC} = 1.6 \times 10^{12} \left(\frac{L_{sync}^{2}}{cR^{2}\nu_{sync}^{4}
\delta^{2} (1+z)^2} \right) \nu_{SSC}^{2} \quad {\rm erg~s^{-1}} 
\label{eqn5}
\end{equation} 
where $L$, $R$, $c$, and $\nu$ are in the same units as in Eqs 3 \& 4.  
The VLBI data and $\gamma$-ray opacity argument 
suggest that $5 < \delta < 20$ for most blazars (e.g.,\cite{vermeulen}; \cite{dondi95}).
This corresponds to the lines (c) and (d) in Fig. 3, respectively 
for $\delta$=5 and 20.  
The above four constraints correspond to the shaded area of Fig. 3, where 
for illustration, we use the overall spectral energy distribution for the 
QHB CTA~102.  

Since we have to use a unique value in calculating the physical parameters,
we use $\delta$ = 10 as a geometrical mean between 5 and 20.  The 
$L_{SSC}$ - $\nu_{SSC}$ line calculated from Eq. 5 corresponding to 
$\delta$ = 10 intersects both the extrapolation of the {\sl ASCA} spectrum 
and the highest {\sl ASCA} value, and the intersections yield the lower 
and upper values for both $L_{SSC}$ and $\nu_{SSC}$.  We adopt a mean of 
these values, which are given in Table 2, and plotted in Figure 4c. 
We used \ginga data instead of \asca data for 3C~279 
because a simultaneous campaign from radio to $\gamma$--ray bands 
was conducted during \ginga observation. 
The values for the other blazars where $\delta$ derived 
from Eq. \ref{eqn4} is $<20$, 
are calculated by assuming $L_{SSC}$ = $L_{HE}$.  
For LBL AO0235+164 where $\delta$ derived from 
Eq. 4 is $>20$, we assume $L_{SSC}$ = $L_{LE}$ 
because the \asca spectra of AO0235+164 is thought 
to be mixture of the LE and HE component, 
as discussed by \cite{madejski96} based on the \rosat and \asca spectra, 
so that the above method may be inappropriate.
For two QHBs (3C~273, PKS~0208-512) the lower limits of $\delta$ are $\sim$5.
The fact that observed $\gamma$-ray flux of PKS~0208-512 
is much higher than the extrapolation of \asca spectrum implies 
the $\gamma$-ray peak is not solely due to SSC mechanism.
Therefore we applied the above method to this source.
On the other hand, since the $\gamma$-ray spectrum of 3C~273 
is below the extrapolation of \asca spectrum, 
the $\gamma$-ray emission is assumed to be due to SSC mechanism  
so there, we assume $L_{SSC}$ = $L_{HE}$.  It is important to note, 
however, that 3C~273 is unique as compared to other blazars considered here 
in that the ``blue bump'' is very pronounced, and thus it is unlikely 
that the jet dominates the entire electromagnetic emission, and therefore, 
a more complex analysis is necessary (see, e.g., \cite{montigny97} for further 
discussion).  

Once we obtain $L_{SSC}$, and $\nu_{SSC}$, 
we can calculate the strength of the magnetic field and the electron Lorentz 
factor $\gamma_{b}$ from Eq. (\ref{eqn1}), (\ref{eqn2}) and those 
are given as follows:
\begin{equation}
B = 0.27 \left(\frac{R_{\mbox{pc}}}{10^{-2}}\right)^{-1}
\left(\frac{\delta}{10}\right)^{-2} 
\sqrt{\left(\frac{L_{sync}}{10^{46}}\right)
\left(\frac{L_{sync}}{L_{SSC}}\right)} \quad {\rm Gauss}
\end{equation}
\begin{equation}
\gamma_{b} = 1.8\times 10^3
\left(\frac{R_{\mbox{pc}}}{10^{-2}}\right)^{1/2}
\left(\frac{\delta}{10}\right)^{1/2} 
\left(\frac{\nu_{sync}(1+z)}{10^{13}}\right)^{1/2} 
\left[\left(\frac{L_{sync}}{10^{46}}\right)
\left(\frac{L_{sync}}{L_{ssc}}\right)\right]^{-1/4}
\end{equation} 
where $R_{\mbox{pc}}$ is size of emission region in parsecs, and other 
quantities are as in Eqs 3, 4, \& 5.  
As before, the upper limit of the size $R$ can be estimated from  
the observed time variability ($\Delta t$) from Table 2, given by 
$R \lesssim c\Delta t \delta / (1+z)$.  

Our calculated values of 
$B$ and $\gamma_{b}$ are plotted respectively in Figures 4d
and 4e.  In these Figures, we also plot the values calculated with 
$R = 0.01$ pc, which would correspond to an observed variability time 
scale of $\sim 1$ day.  

From our analysis, the magnetic field for blazars observed with \asca 
is inferred to be 0.1 -- 1 Gauss. 
The value of $B$ is comparable between the different subclasses of blazars, 
although $B$ is somewhat lower in HBLs than in QHBs. With these values of $B$, 
we estimate $\gamma_{b}$ to be $10^{3} - 10^{4}$ for QHBs, and
$10^{5}$ for HBLs.  The differences of $\gamma_{b}$  between different 
sub-classes of blazars imply that the relativistic electrons 
are accelerated to higher energies in HBLs than in QHBs. 
Alternatively, higher $\gamma_{b}$ in HBLs might be obtained by 
increasing $\delta$.  However, in those objects, we believe that there is 
no contribution to the $\gamma$--ray production from other mechanisms 
besides SSC, and thus the observed $L_{HE}$ is $L_{SSC}$.  In such case, 
$\gamma_{b}$ depends on $\delta$ only linearily 
(cf. Eq. 7 and $R\lesssim c\Delta t \delta/(1+z)$), 
and thus varying $\delta$ to be 5 or 20 respectively decreases or 
increases our derived $\gamma_{b}$ only by a factor of two, which is small
when compared to the large difference of $\gamma_{b}$ calculated by us (cf. 
Fig. 4e).  

In QHBs the strong optical and UV line emission implies a presence of 
dense external radiation fields.  This means that in the frame of 
reference of the jet, these can easily dominate over the internal 
synchrotron radiation, resulting in the ERC emission dominating over 
the SSC emission in $\gamma$--ray band (e.g., \cite{sikora97}).  
It is likely that the difference of $\gamma_{b}$ is most likely due 
to the large photon density in QHBs as compared with that of HBLs. It should
be noted that TeV photons have been observed 
only from HBLs, where we calculate higher values of $\gamma_{b}$. 

\section{Summary}

The non-thermal emission from blazars, observed from radio to GeV/TeV 
$\gamma$--rays, is thought to be the result of radiation of very energetic 
particles via both synchrotron and Compton processes.  The overall 
spectra of all blazars over this wide range of energies appear similar 
from one object to another, forming two distinct peaks in the $\nu F(\nu)$ 
representation.  The X--ray regime is important, as it is where the 
emission due to both processes overlaps:  for the High-energy peaked BL Lac
objects (HBLs), X--rays form the high energy tail of the synchrotron 
emission, while for blazars showing quasar-like emission lines (QHBs), they 
form the lowest observable energy end of the Comptonized spectrum.  
The X--ray spectra of Low-energy peaked BL Lac objects (LBLs) appear 
intermediate, which may depend on the level of activity of the object.  
An application of the synchrotron self-Compton (SSC) model to the overall 
spectral distribution and variability data of 18 blazars observed by \asca 
implies that for the HBLs, the SSC model can explain all available data 
quite well, and implies relativistic Doppler factors $\delta$ in the range 
of 5 - 20, consistent with the those derived from the VLBI data and from 
the limits inferred from $\gamma$--ray opacity to pair production, 
$\gamma\gamma\rightarrow e^{+}e^{-}$.  Further support for it comes from 
the good agreement of values physical parameters of the radiating plasma 
as inferred from the energetics requirement to produce the TeV $\gamma$--rays, 
with that inferred from the spectral variability observed in X--rays in 
a number of objects.  

The situation in QHBs (and in some LBLs) is quite different, as the 
application of the SSC model implies the values of $\delta$ much in excess 
of values inferred from the $\gamma$--ray opacity arguments, or the 
VLBI data.  This discrepancy can be eliminated if we adopt a scenario 
where the observed GeV $\gamma$--rays are produced by Comptonization 
of external photons (via the ERC process), 
while the flux produced by the SSC process -- while present and possibly 
responsible for the X--ray emission -- in the GeV range is well below the 
ERC emission. With this we estimated the $L_{SSC}$ of QHBs, 
by assuming X--ray emission in \asca band is due to SSC process.

The inferred magnetic fields in all blazars is comparable 
-- $\sim$ 0.1 to 1 Gauss -- but, what distinguishes 
QHBs and HBLs is the much lower value of Lorentz factors of electrons 
radiating at the peak, $\sim 10^{3}$ -- $10^{4}$ for QHBs against $\sim 10^5$ 
for HBLs.  However, the details of the jet structure, and in particular, 
the question of the particle acceleration, still remain open, with the 
hope that the future missions -- such as the upcoming satellites
{\sl ASTRO-E} and {\sl GLAST} -- 
will further advance our knowledge of the details of 
physical processes in these extreme, enigmatic objects.  

\acknowledgments
This research has made use of NASA/IPAC Extragalactic Database (NED) and data 
from the University of Michigan Radio Astronomy Observatory (UMRAO).
We thank H. Ter\"asranta and M. Tornikoski for allowing us to use 
their unpublished radio data;  M. Sikora for many useful discussions 
and comments; collaborators of \asca observation, 
W. Collmar, J. G. Stacy, S. Sembay, A. Yamashita, E. Idesawa, and J. Kataoka;  
and the referee, C. M. Urry, for valuable comments.
G. M acknowledges the support of NASA grant NAG5-4106
and of the Grants-in Aid for Scientific Research
by Ministry of Education, Culture, and Science (Monbusho) of Japan (08044103).

\begin{deluxetable}{llrcccc}
\footnotesize
\tablecaption{\asca observation of blazars.\label{tbl-1}}
\tablewidth{0pt}
\tablehead{
\colhead{source} & \colhead{other name} & 
\colhead{redshift}  & \colhead{$N^{\mbox{\footnotesize Gal}}_{\mbox{\small H}}$~\tablenotemark{a}} & \colhead{date\tablenotemark{b}} & \colhead{$\alpha_X$\tablenotemark{c}} & \colhead{flux$\tablenotemark{d}$}} 
\startdata
\multicolumn{7}{c}{HBL} \nl \hline 
0323+022 & H   & 0.147 & 8.68 & 94 Jan 24 & 1.74$^{+0.19}_{-0.13}$ & (1.0$\pm$0.1)$\times 10^{-12}$ \nl
0414+009 & H   & 0.287 & 9.15 & 96 Aug 30 & 1.62$^{+0.04}_{-0.05}$ & (7.4$\pm$0.1)$\times 10^{-12}$ \nl
0548--322 & PKS & 0.069 & 2.49 & 93 Oct 30 & 0.96$^{+0.02}_{-0.02}$ & (3.4$\pm$0.1)$\times 10^{-11}$ \nl
1101+384 & Mkn~421 & 0.031 & 1.45 & 93 May 10 & 1.97$^{+0.02}_{-0.03}$ & (2.4$\pm$0.1)$\times 10^{-11}$ \nl
1426+428 & H   & 0.129 & 1.37 & 94 Feb 6 & 1.17$^{+0.04}_{-0.03}$ & (1.8$\pm$0.1)$\times 10^{-11}$ \nl
1652+398 & Mkn~501 & 0.034 & 1.73 & $^{\ast}$96 Mar 26, 27 & 1.22$^{+0.02}_{-0.02}$ & (6.6$\pm$0.1)$\times 10^{-11}$ \nl 
2155$-$304 & PKS & 0.116 & 1.77 & 94 May 19 & 1.62$^{+0.01}_{-0.01}$ & (7.2$\pm$0.1)$\times 10^{-11}$ \nl 
\cutinhead{LBL}
0235+164 & AO & 0.940 & 7.60 & $^{\ast}$94 Feb 4,11,16,19 & 0.82$^{+0.14}_{-0.12}$ & (1.4$\pm$0.1)$\times 10^{-12}$ \nl
0716+714 &    & $>$0.3 & 3.73 & $^{\ast}$94 Mar 16,19,21 & 1.07$^{+0.10}_{-0.14}$ & (1.3$\pm$0.1)$\times 10^{-12}$ \nl
0735+178 & PKS & $>$0.424 & 4.35 & 96 Apr 22 & 0.76$^{+0.18}_{-0.23}$ & (7.8$\pm$0.7)$\times 10^{-13}$ \nl
0851+202 & OJ287 & 0.306 & 2.75 & 94 Nov 18 & 0.62$^{+0.05}_{-0.06}$ & (5.1$\pm$0.1)$\times 10^{-12}$ \nl
\cutinhead{QHB} 
0208$-$512 & PKS & 1.003 & 3.17 & $^{\ast}$95 Jan 15 & 0.66$^{+0.10}_{-0.09}$ & (7.9$\pm$0.3)$\times 10^{-12}$ \nl
0333+321 & NRAO140 & 1.258 & 14.22 & 94 Feb 1 & 0.60$^{+0.04}_{-0.05}$ & (9.7$\pm$0.2)$\times 10^{-12}$ \nl
0528+134 & PKS & 2.060 & 25.30 & $^{\ast}$95 Mar 7,14,19 & 0.58$^{+0.05}_{-0.04}$ & (1.1$\pm$0.1)$\times 10^{-11}$ \nl
1226+023 & 3C~273 & 0.158 & 1.81 & $^{\ast}$93 Dec 20,23,27 & 0.51$^{+0.01}_{-0.01}$ & (1.7$\pm$0.1)$\times 10^{-10}$ \nl
1252--055 & 3C~279 & 0.538 & 2.22 & $^{\ast}$93 Dec 21,23,27 & 0.65$^{+0.06}_{-0.07}$ & (1.0$\pm$0.1)$\times 10^{-11}$ \nl
1633+382 & 4C~38.41 & 1.814 & 1.00 & $^{\ast}$96 Mar 21,25,27 & 0.51$^{+0.32}_{-0.31}$ & (2.1$\pm$0.3)$\times 10^{-12}$ \nl
2230+114 & CTA~102 & 1.037 & 5.05 & $^{\ast}$95 Dec 6 & 0.58$^{+0.12}_{-0.12}$ & (3.1$\pm$0.2)$\times 10^{-12}$ \nl
\tablenotetext{a}{Galactic column density in the units of 10$^{20}$ cm$^{-2}$.
\protect{\cite{elvis89}} when available, \protect{\cite{stark92}} otherwise. For PKS~0208--512 the value is from \protect{\cite{dickey90}}.}
\tablenotetext{b}{$\ast$ symbol show simultaneous observation with \asca and EGRET.}
\tablenotetext{c}{2--10 keV energy index obtained with the absorption fixed at Galactic value.}
\tablenotetext{d}{Absorbed 2--10 keV flux in unit of erg~s$^{-1}$ cm$^{-2}$. 
Systematic errors 10 \% are not included.}
\tablenotetext{}{Errors are quoted at 90\% confidence limit for single parameter.}
\enddata
\end{deluxetable}

\clearpage

\begin{deluxetable}{lcccccccccccc}
\scriptsize
\tablecaption{Results of multiband analysis.\label{tbl-2}}
\tablewidth{0pt}
\tablehead{
\colhead{source} & 
\multicolumn{3}{c}{log frequency(Hz)\tablenotemark{a}} & \colhead{} & 
\multicolumn{3}{c}{log luminosity(erg~s$^{-1}$)\tablenotemark{b}} & \colhead{} 
& \multicolumn{3}{c}{$\Delta t$} & 
\colhead{$\delta$\tablenotemark{c}} \\
\cline{2-4} \cline{6-8} \cline{10-12} \\
\colhead{} & \colhead{$\nu_{LE}$\tablenotemark{d}} & \colhead{$\nu_{HE}$} & \colhead{$\nu_{SSC}$} & 
\colhead{} & 
\colhead{$L_{LE}$\tablenotemark{d}} & \colhead{$L_{HE}$\tablenotemark{d}} & 
\colhead{$L_{SSC}$} & \colhead{} & \colhead{(days)} & \colhead{band} & ref & \colhead{}} 
\startdata
H0323+022 & 15.5 & \nodata & \nodata &  & 44.4 & $<$44.6 & $<$44.6 & & 
\nodata & \nodata & \nodata & \nodata \nl
H0414+009 & 16.8 & \nodata & \nodata &  & 45.3 & $<$45.4 & $<$45.4 & &
\nodata & \nodata & \nodata & \nodata \nl
PKS0548--322 & 16.8 & \nodata & \nodata & & 44.4 & $<$44.1 & $<$44.1 
& & \nodata & \nodata & \nodata & \nodata \nl
Mkn~421 & 16.0 & 24.9$\pm$0.3 & 24.9$\pm$0.3 & & 44.3 & 43.9 & 43.9$\pm$0.1 
& & 0.5 & X & 1 & 3 \nl
H1426+428 & 17.0 & \nodata & \nodata & & 44.7 & $<$44.6 & $<$44.6 & &  
\nodata & \nodata & \nodata & \nodata \nl
Mkn~501 & 16.3 & 24.2$\pm$0.6 & 24.2$\pm$0.6 & & 44.2 & 44.3 & 44.3$\pm$0.4 
& & 1 & $\gamma$ & 2 & 1 \nl
PKS2155$-$304 & 16.2 & $>$23.9 & $>$23.9 & & 45.8 & $>$45.2 & $>$45.2 & 
& 0.5 & X & 3 & 2 \nl
AO0235+164 & 13.6 & $>$24.2 & \nodata & & 46.9 & $>$46.9 
& 46.9\tablenotemark{e} & & 3 & X & 4 & 6.6$\times$10$^2$ \nl
0716+714 & 14.6 & $>$23.5 & \nodata & & 45.8 & $>$45.8 & $>$45.8 &  
& 2 & X & 5 & 18 \nl
PKS0735+178 & 14.4 & $<$22.3 & \nodata & & 46.3 & $>$46.6 & $>$46.6 &  
& 1 & opt & 6 & 12 \nl
OJ287 & 13.9 & \nodata & \nodata & & 46.2 & $>$46.8 & \nodata & & 
\nodata & \nodata & \nodata & \nodata \nl
PKS0208$-$512 & 13.6  & 21.0$\pm$0.1 & 20.6$^{+1.0}_{-0.7}$ & &  
46.9 & 48.4 & 46.5$^{+0.6}_{-0.4}$ & & 7 & $\gamma$ & 7 & 5 \nl
NRAO140 & 13.7 & \nodata & \nodata & & 46.2 & $<$46.8 & $<$46.8 & &
\nodata & \nodata & \nodata & \nodata \nl
PKS0528+134 & 13.0 & 22.3$\pm$0.2 & 18.7$^{+0.7}_{-0.1}$ & & 47.1 & 48.4 
& 46.8$^{+0.4}_{-0.1}$ & & 2 & $\gamma$ & 8 & 2.5$\times$10$^2$ \nl
3C~273 & 13.5  & 20.2$\pm$0.1 & 20.2$\pm$0.1 & & 45.8 & 46.2 & 46.2$\pm$0.1 
& & 1 & opt & 9 & 5 \nl
3C~279 & 13.7  & 23.6$\pm$0.4 & 20.7$^{+0.9}_{-0.7}$ & & 46.7 & 47.5 
& 46.6$^{+0.5}_{-0.3}$ & & 2 & $\gamma$ & 10 & 1.5$\times$10$^2$ \nl
4C~38.41 & 13.9  & 23.7$\pm$0.2 & 21.1$^{+1.1}_{-0.9}$ & & 46.4 & 48.2 
& 46.7$^{+1.0}_{-0.6}$ & & 2 & $\gamma$ & 11 & 86 \nl
CTA~102 & 13.5 & 21.6$\pm$0.1 & 20.1$^{+0.9}_{-0.8}$ & & 46.4 & 47.9 
& 46.1$^{+0.6}_{-0.4}$ & & 2 & opt & 12 & 21 \nl
\enddata
\tablenotetext{a}{rest-frame peak frequency}
\tablenotetext{b}{observer-frame luminosity at the peak frequency 
assuming 4$\pi$ radiation}
\tablenotetext{c}{lower limit assuming $L_{HE}$ = $L_{SSC}$; see the text}
\tablenotetext{d}{error is $\pm$0.1}
\tablenotetext{e}{assuming $L_{SSC}$ = $L_{LE}$;  see the text}
\tablerefs{(1)\cite{takahashi96}; (2)\cite{aharonian97}; (3)\cite{kubo97}; \cite{urry97}; (4)Madejski et al. 1996; (5)\cite{cappi94}; (6)\cite{xie92}; (7)\cite{bertsch93}; (8)\cite{hunter93b}; (9)\cite{courvoisier88}; (10)\cite{kniffen93}; (11)\cite{mattox93}; (12)\cite{pica88}}
\end{deluxetable}

\clearpage

\plotone{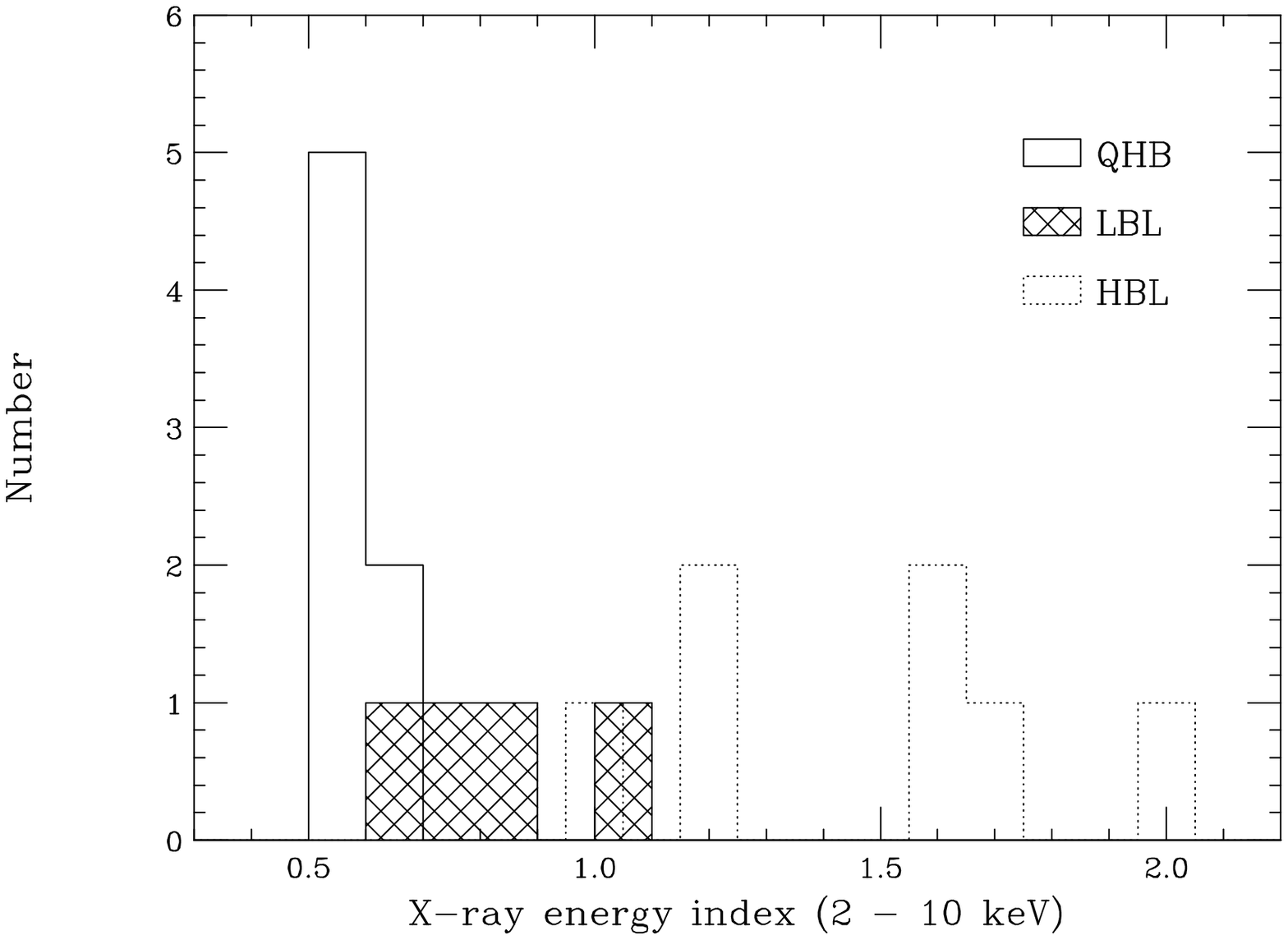}
\figcaption{
Distribution of power law energy indices observed by \asca in the 2--10 keV 
in the observer frame.  The spectra were assumed to be power laws with 
Galactic absorption. \label{fig1}}

\clearpage

\epsscale{0.5}
\plotone{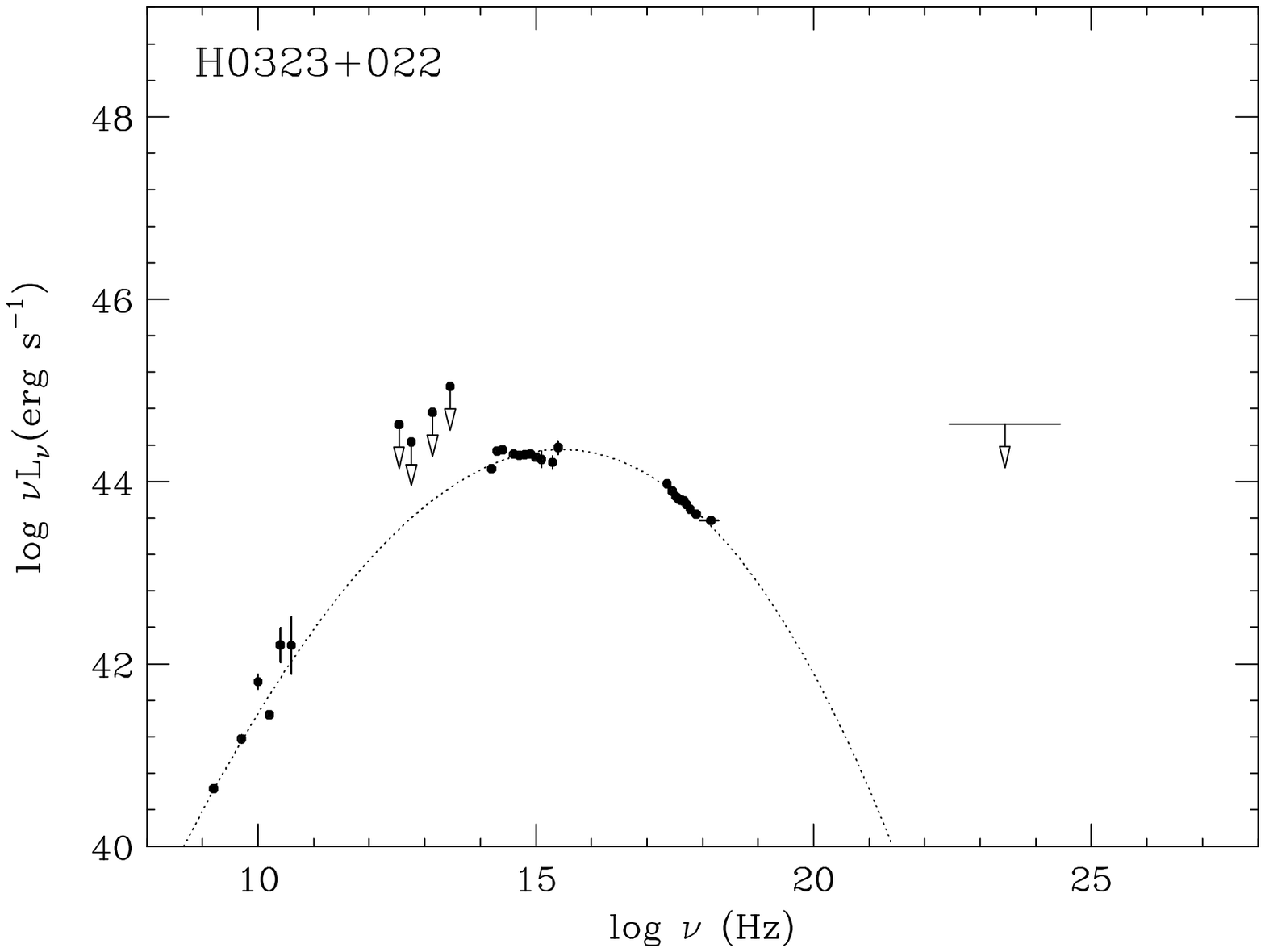}
\plotone{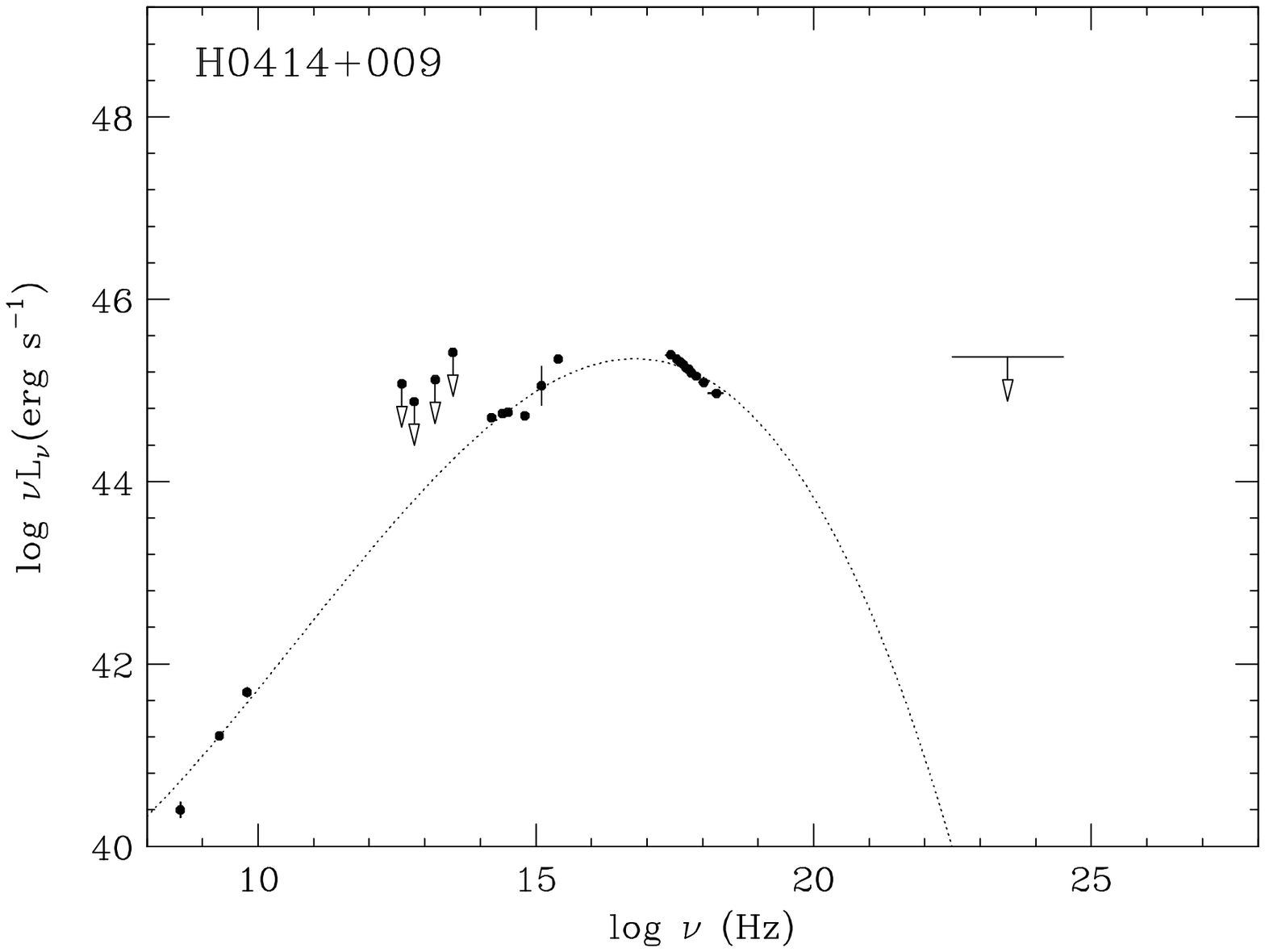}

\plotone{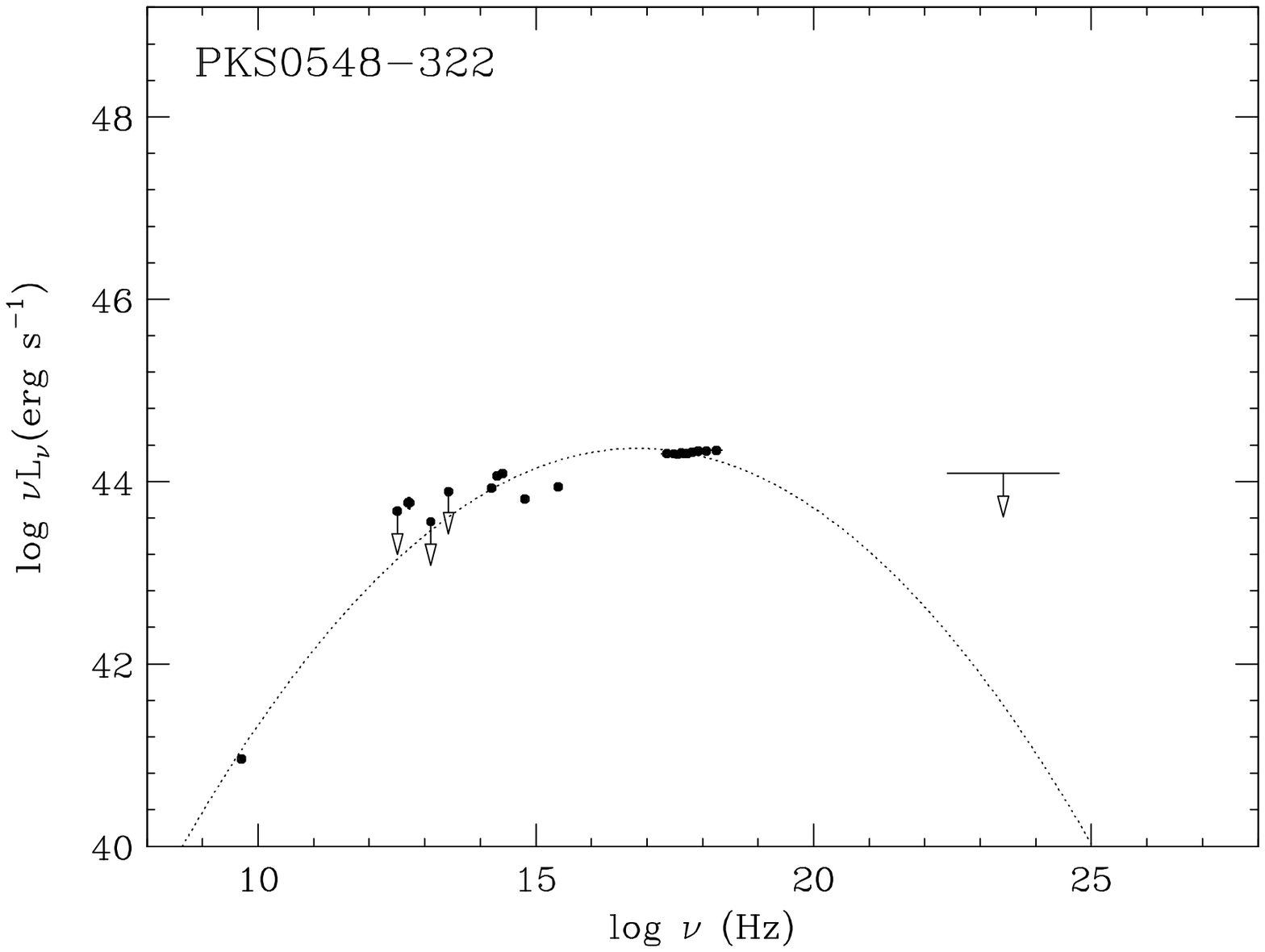}
\plotone{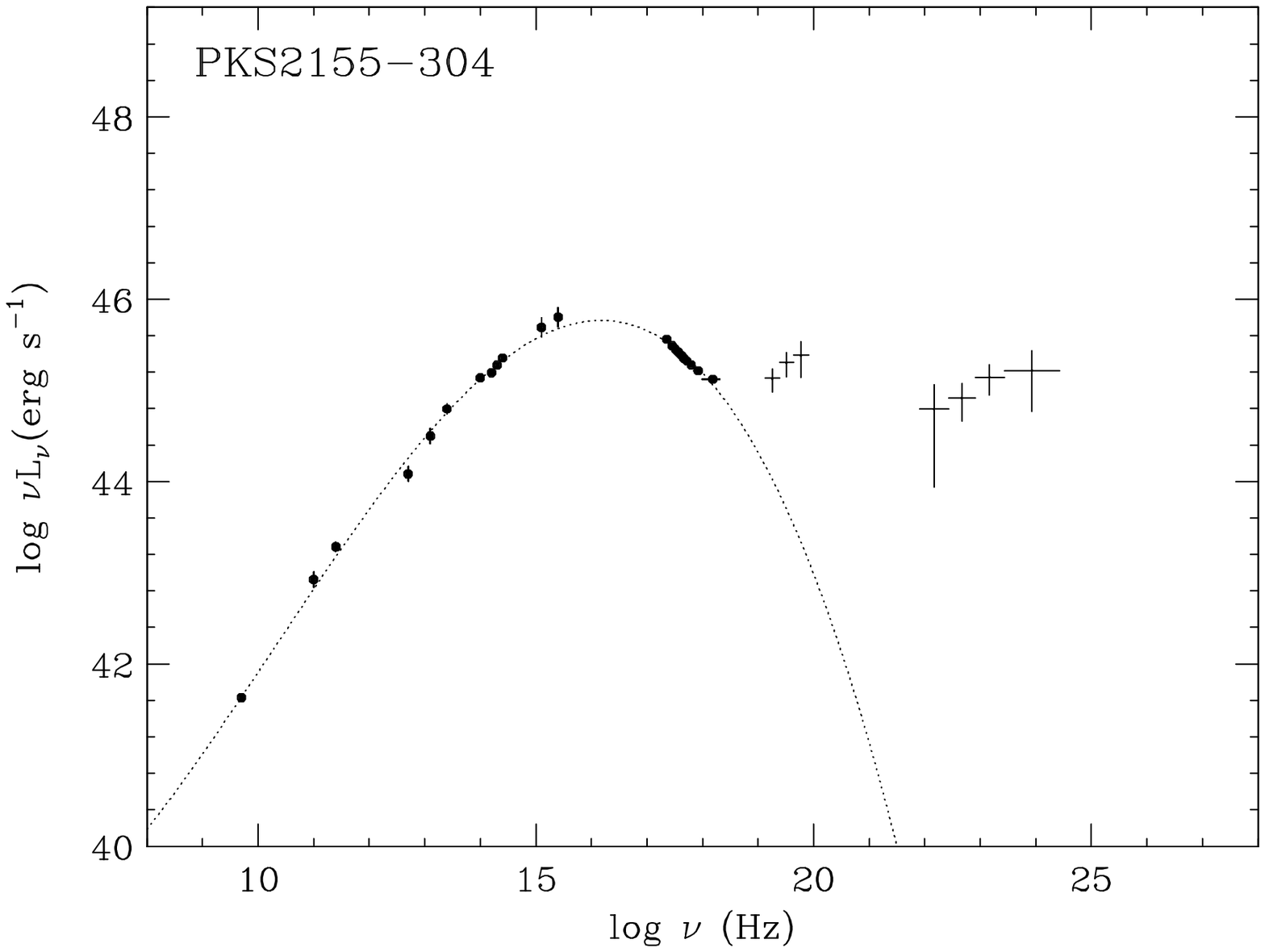}

\plotone{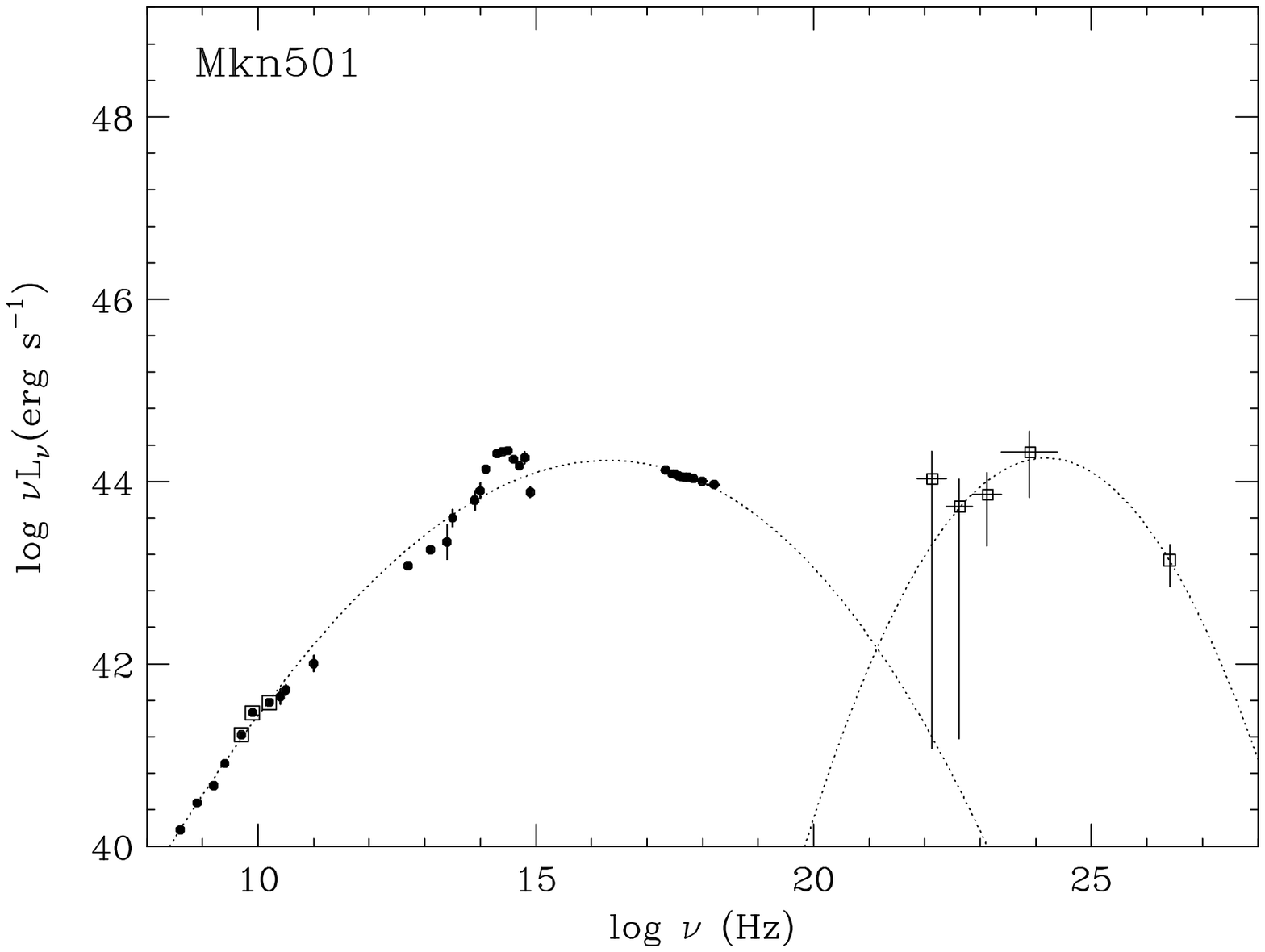}
\plotone{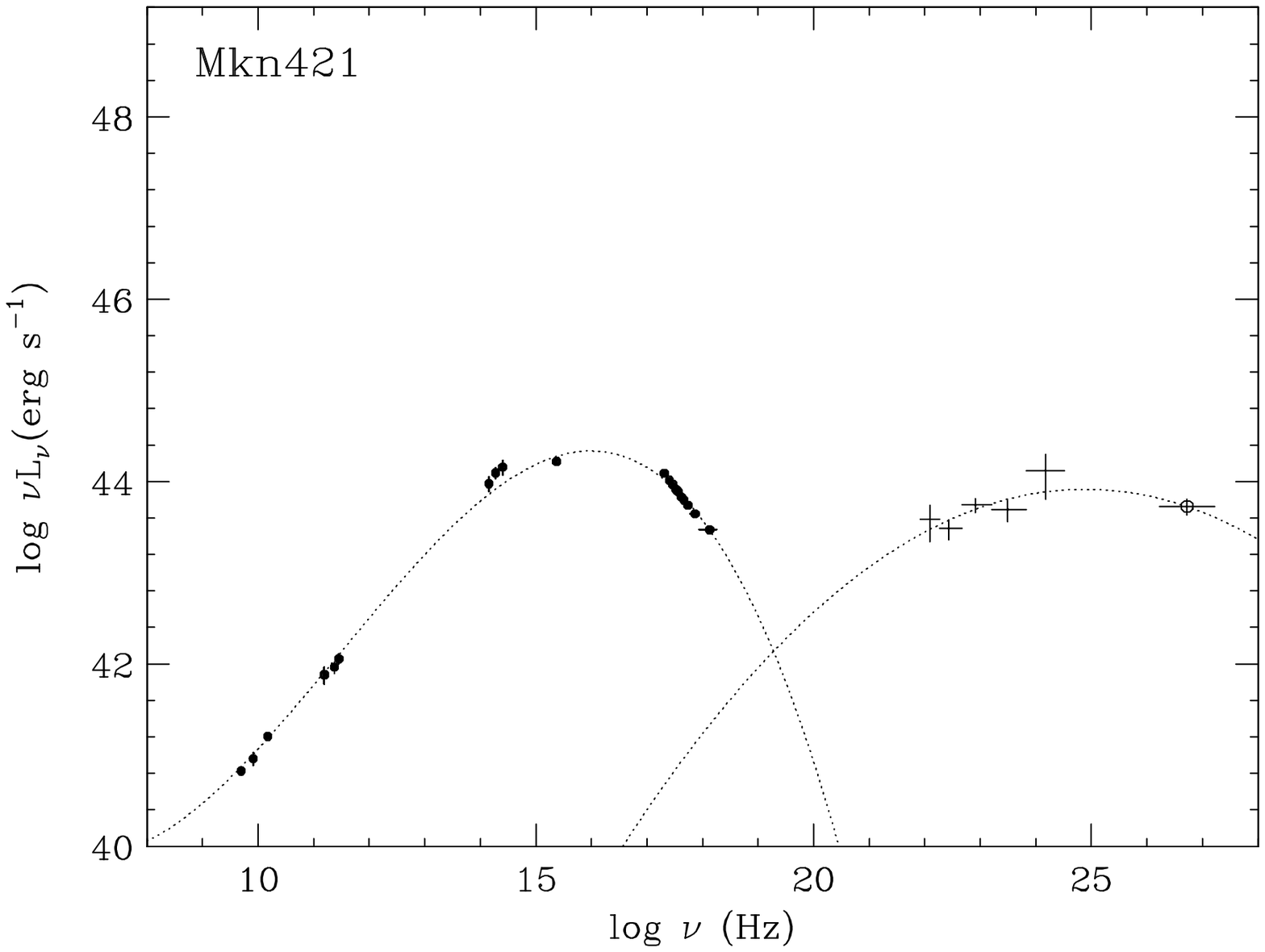}

\clearpage

\plotone{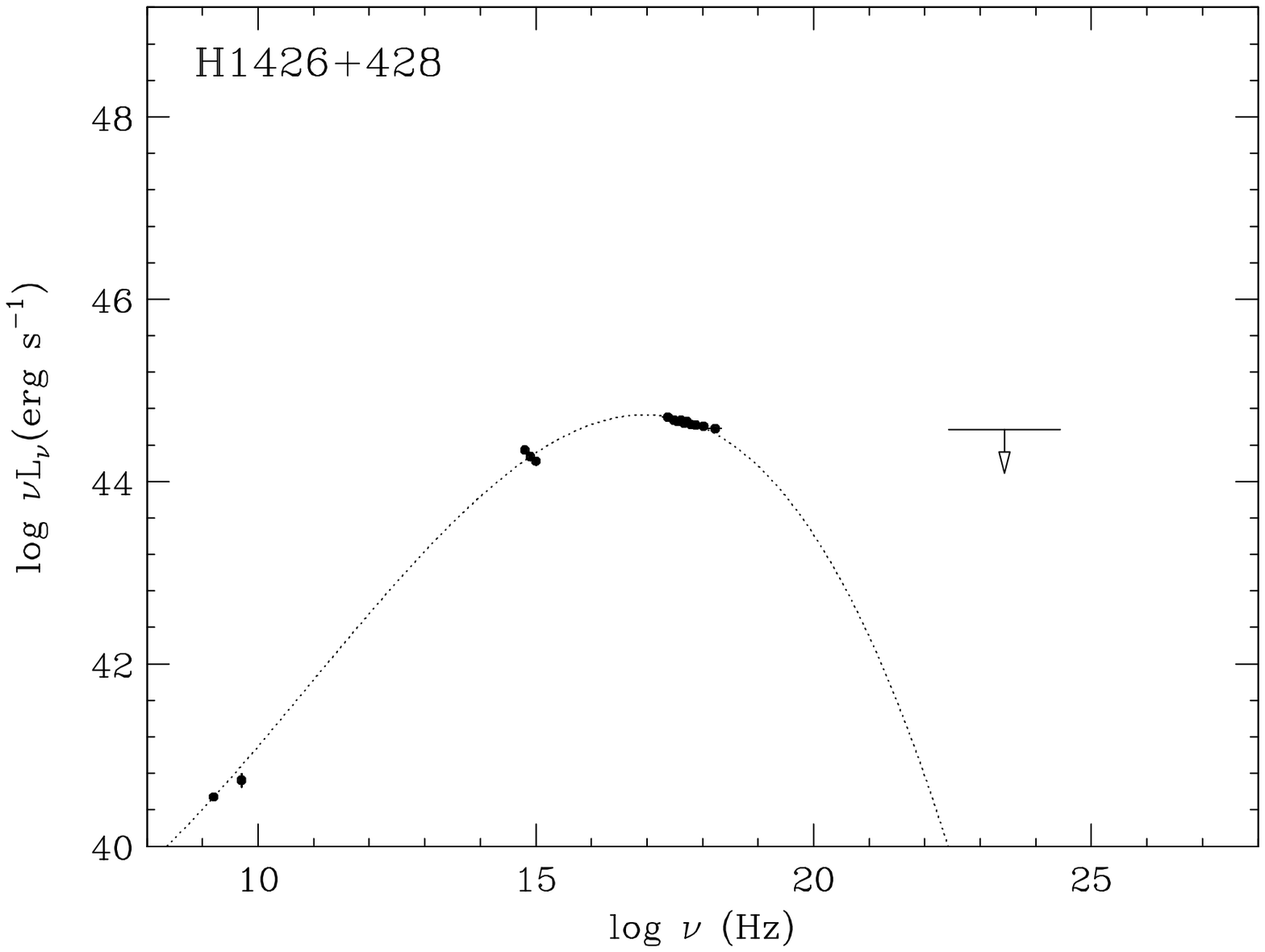}
\plotone{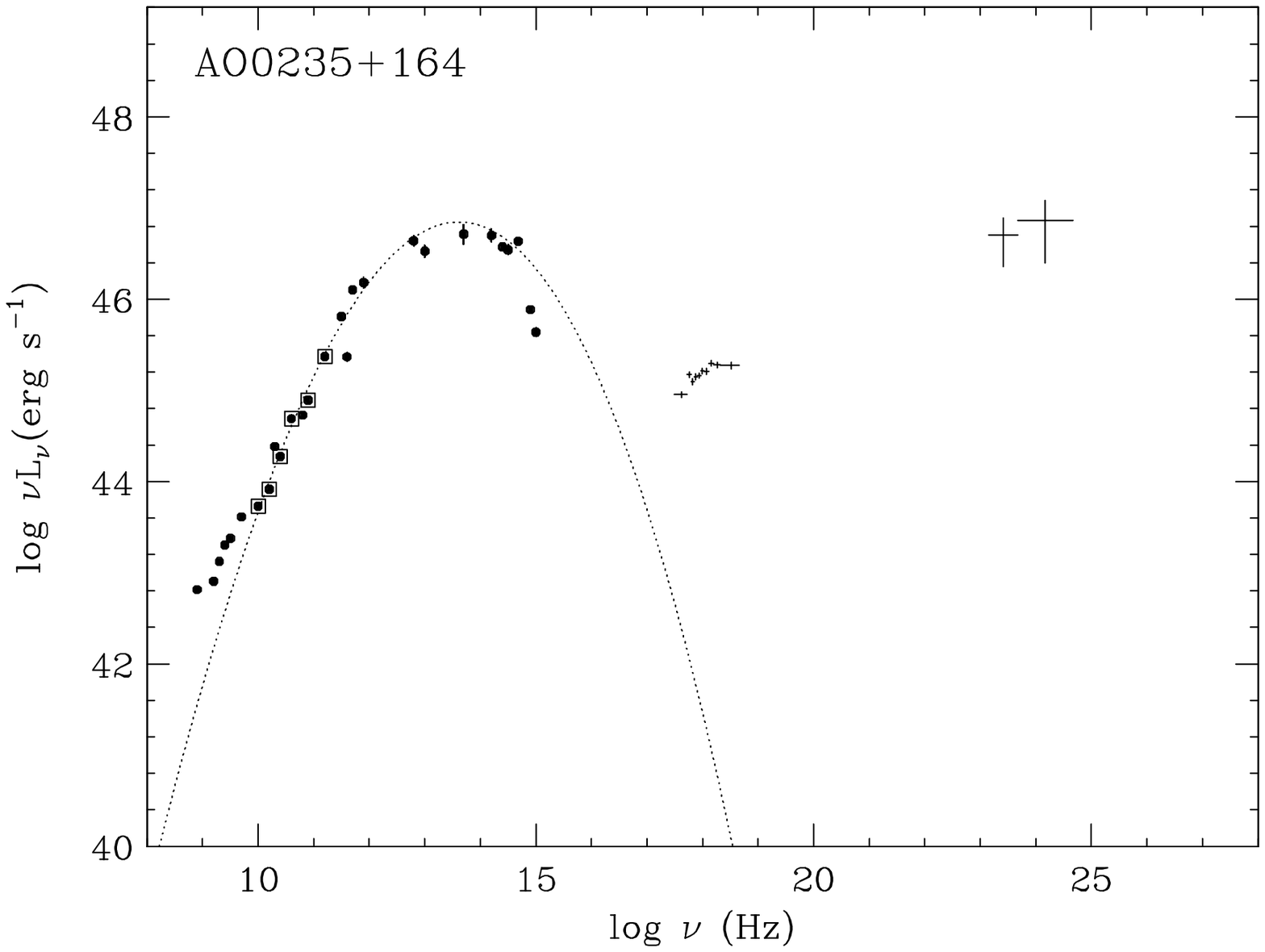}

\plotone{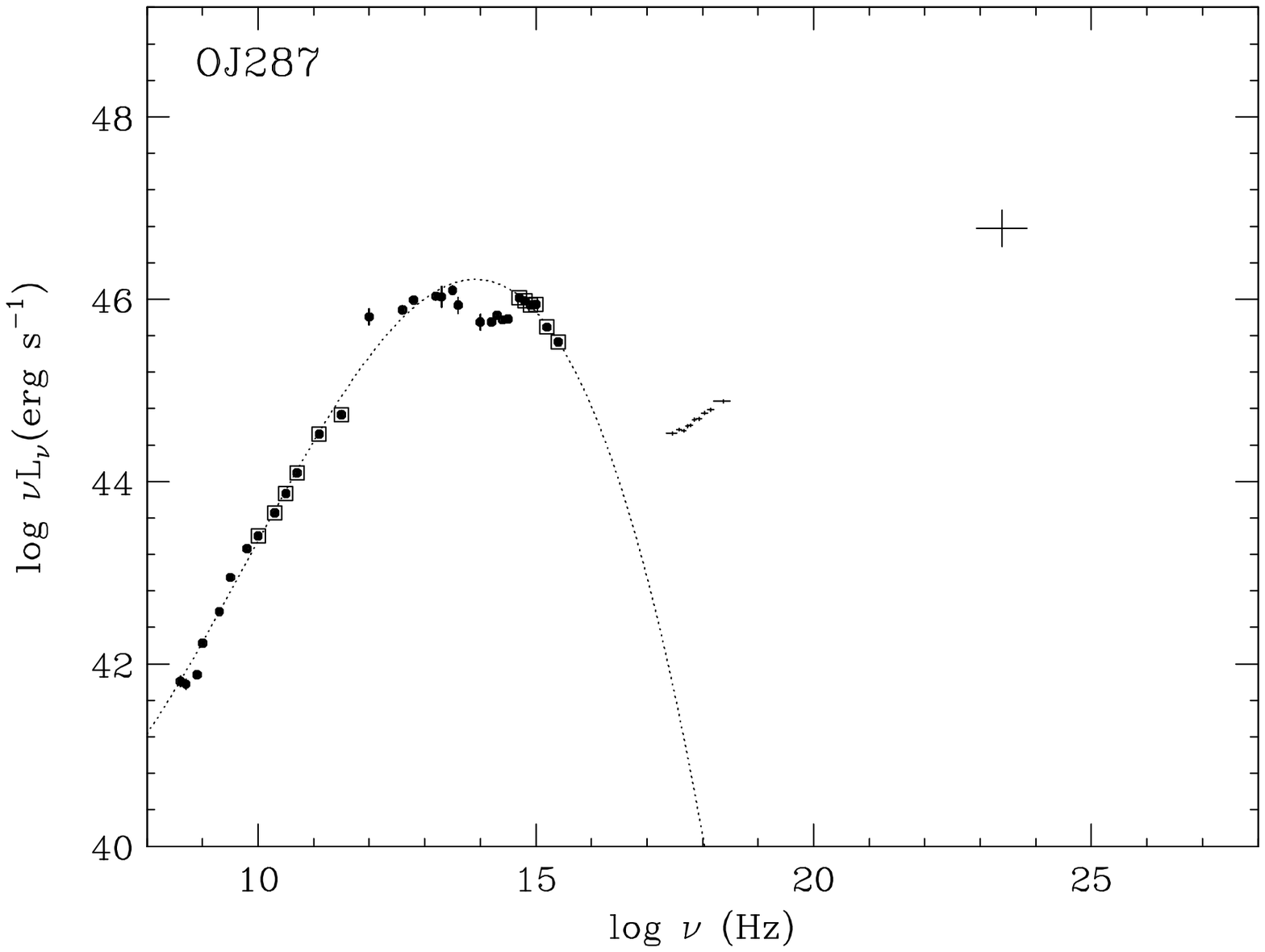}
\plotone{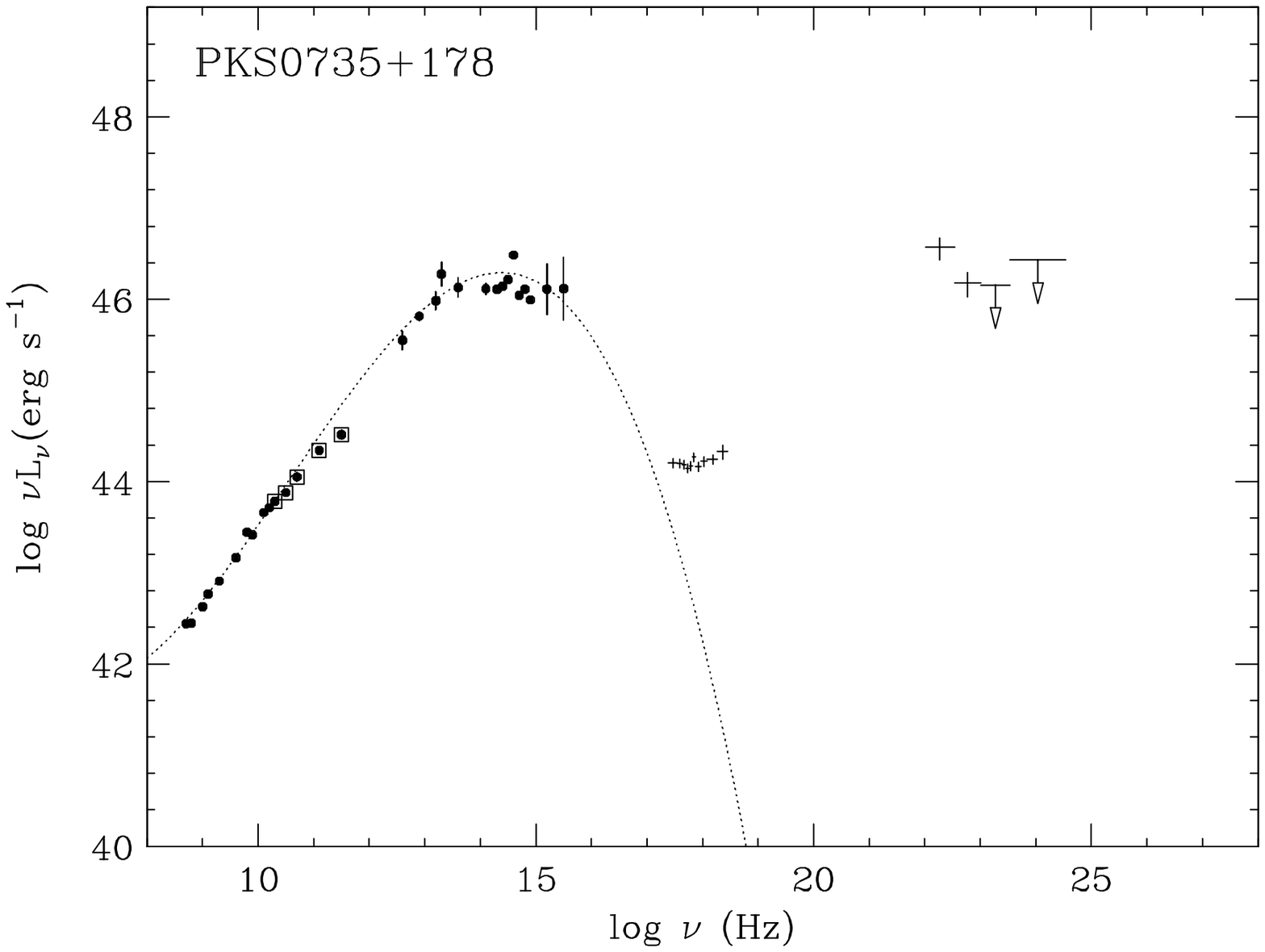}

\plotone{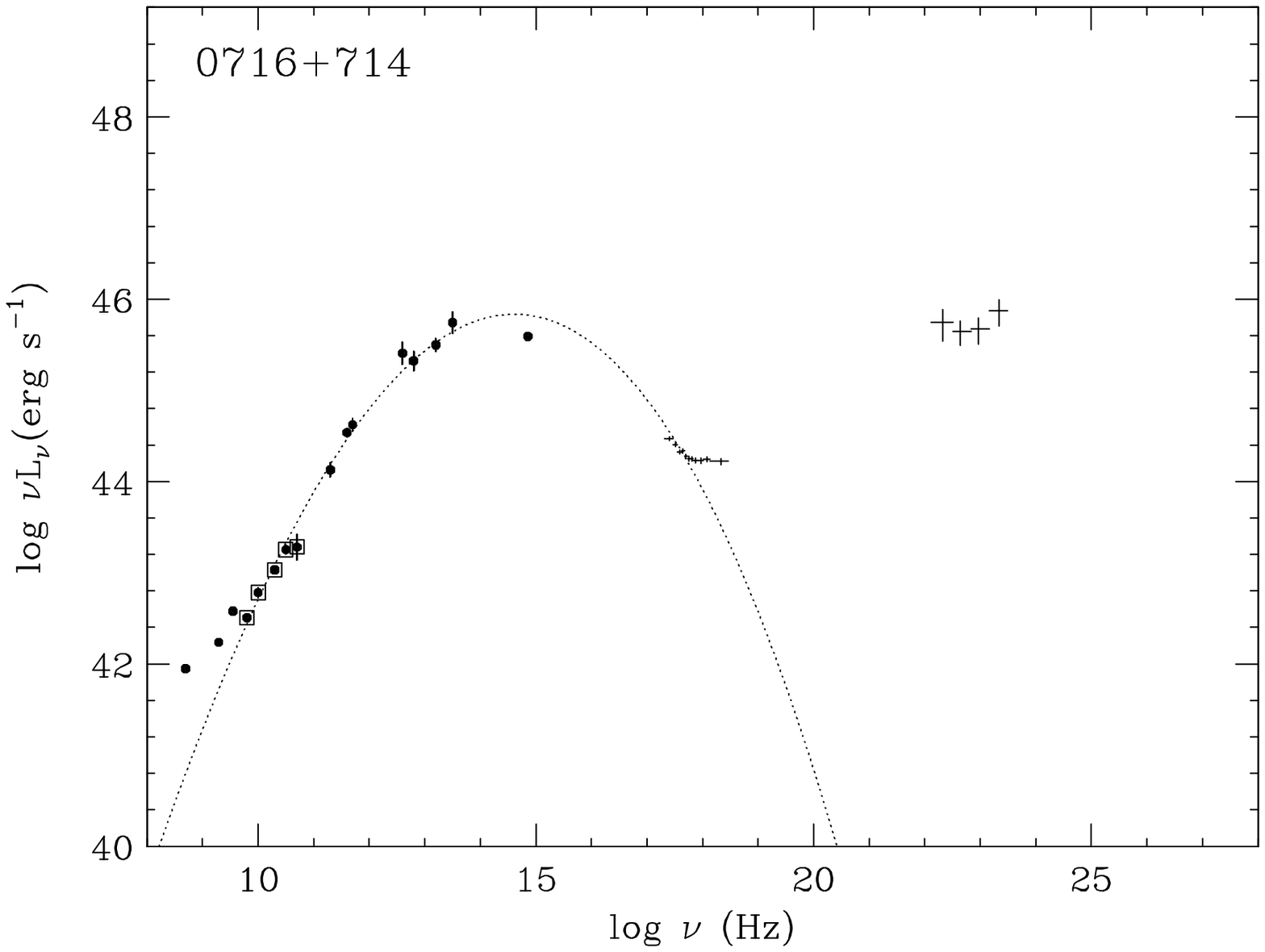}
\plotone{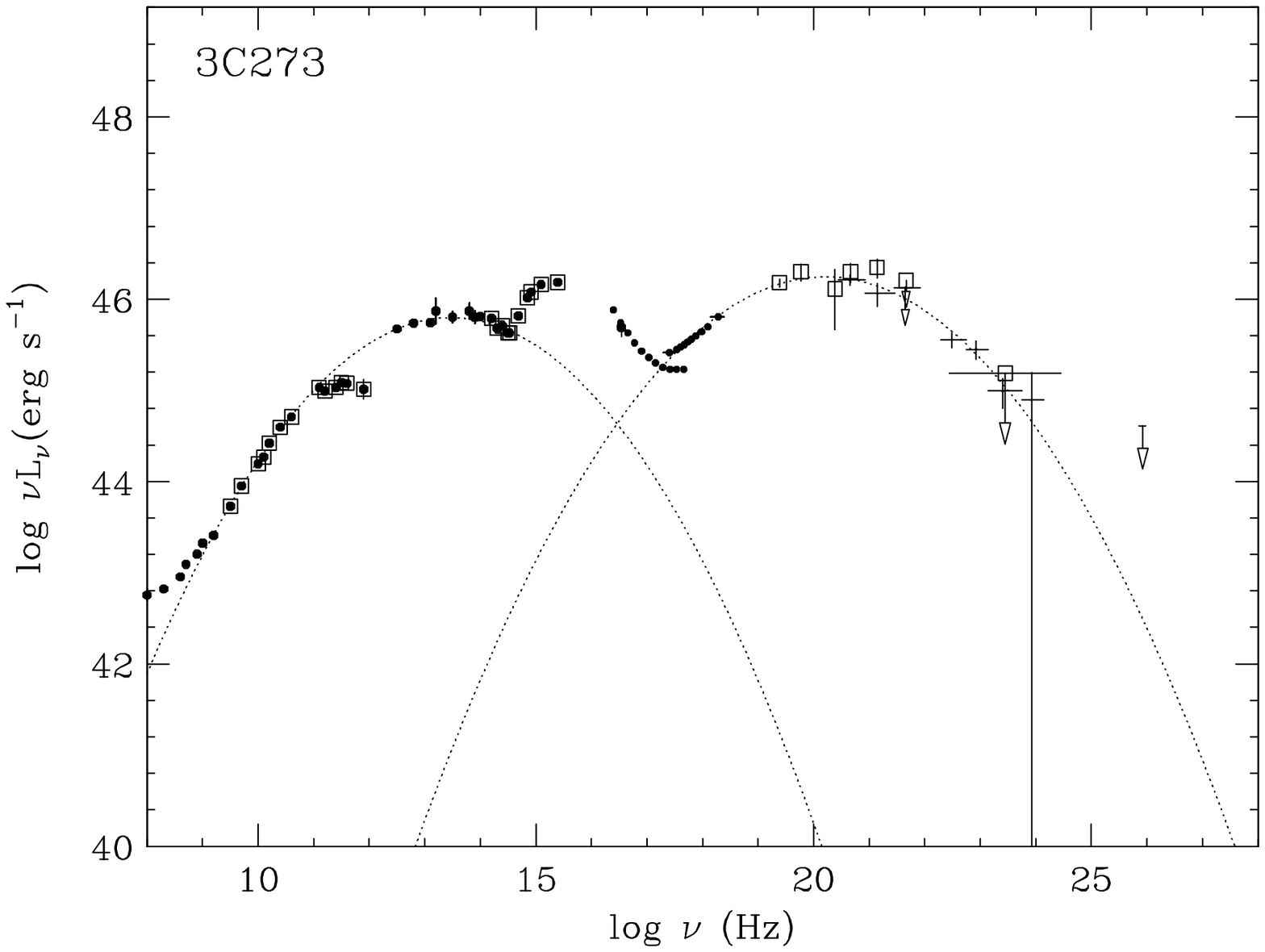}

\clearpage

\plotone{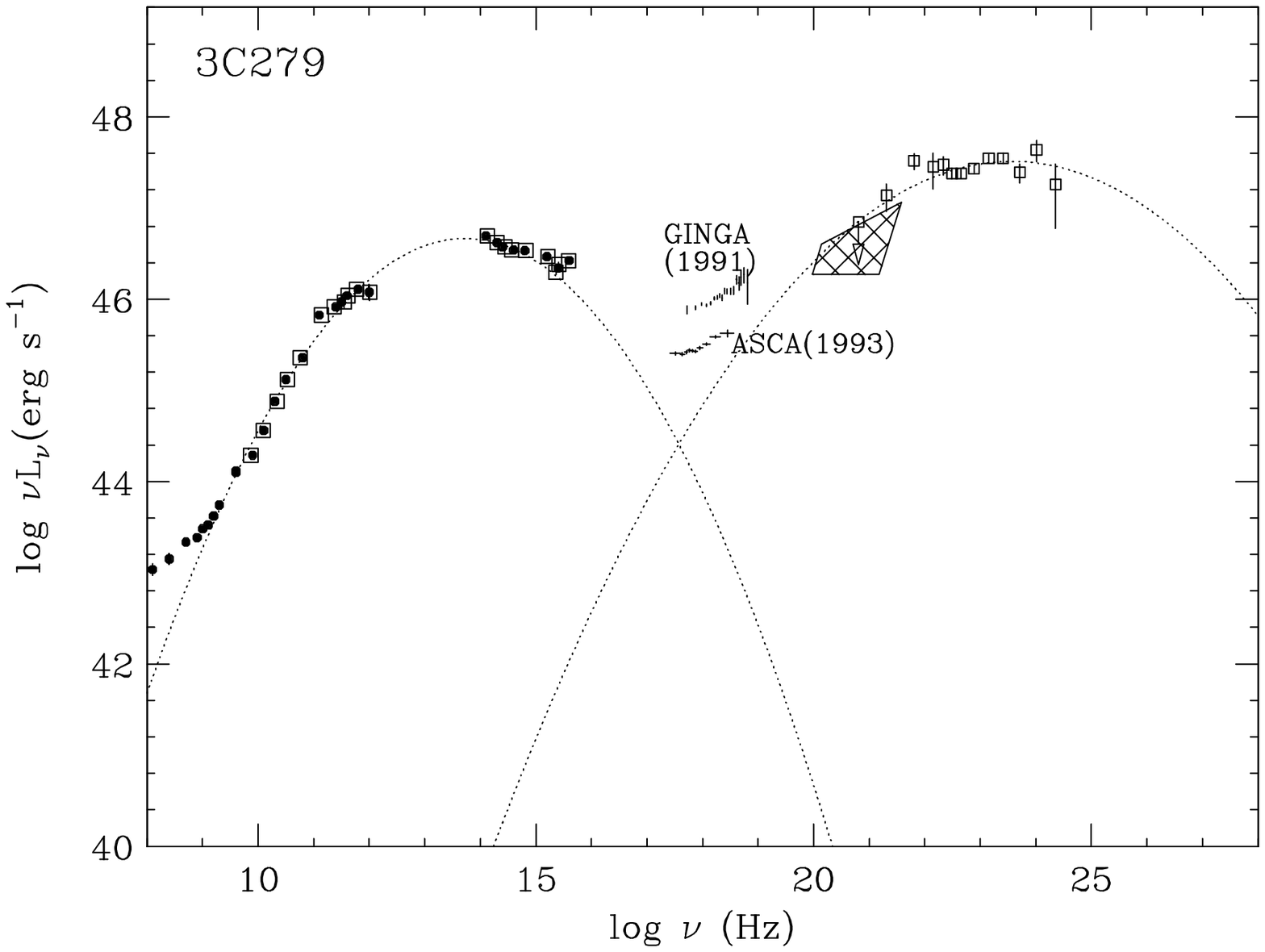}
\plotone{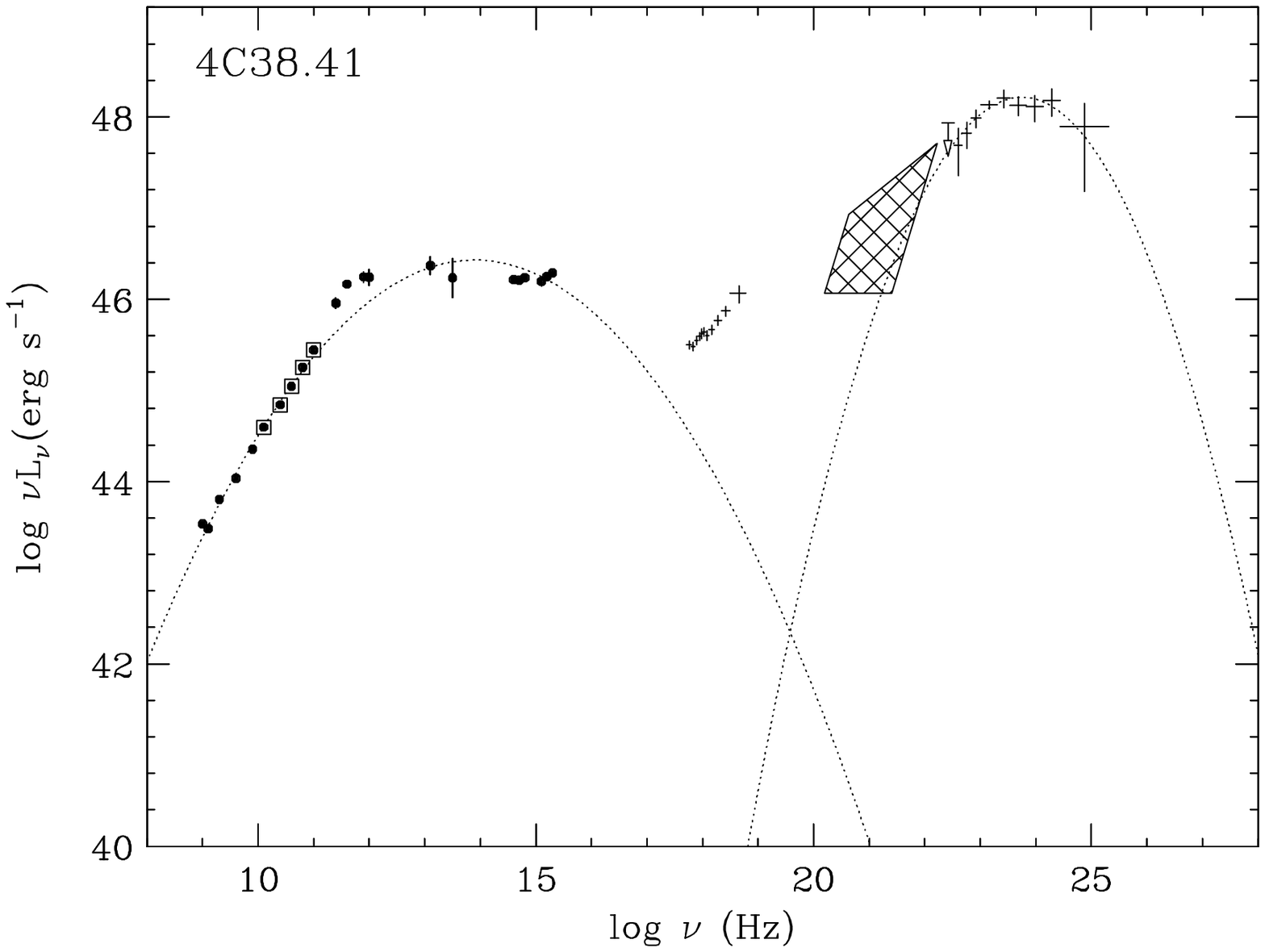}

\plotone{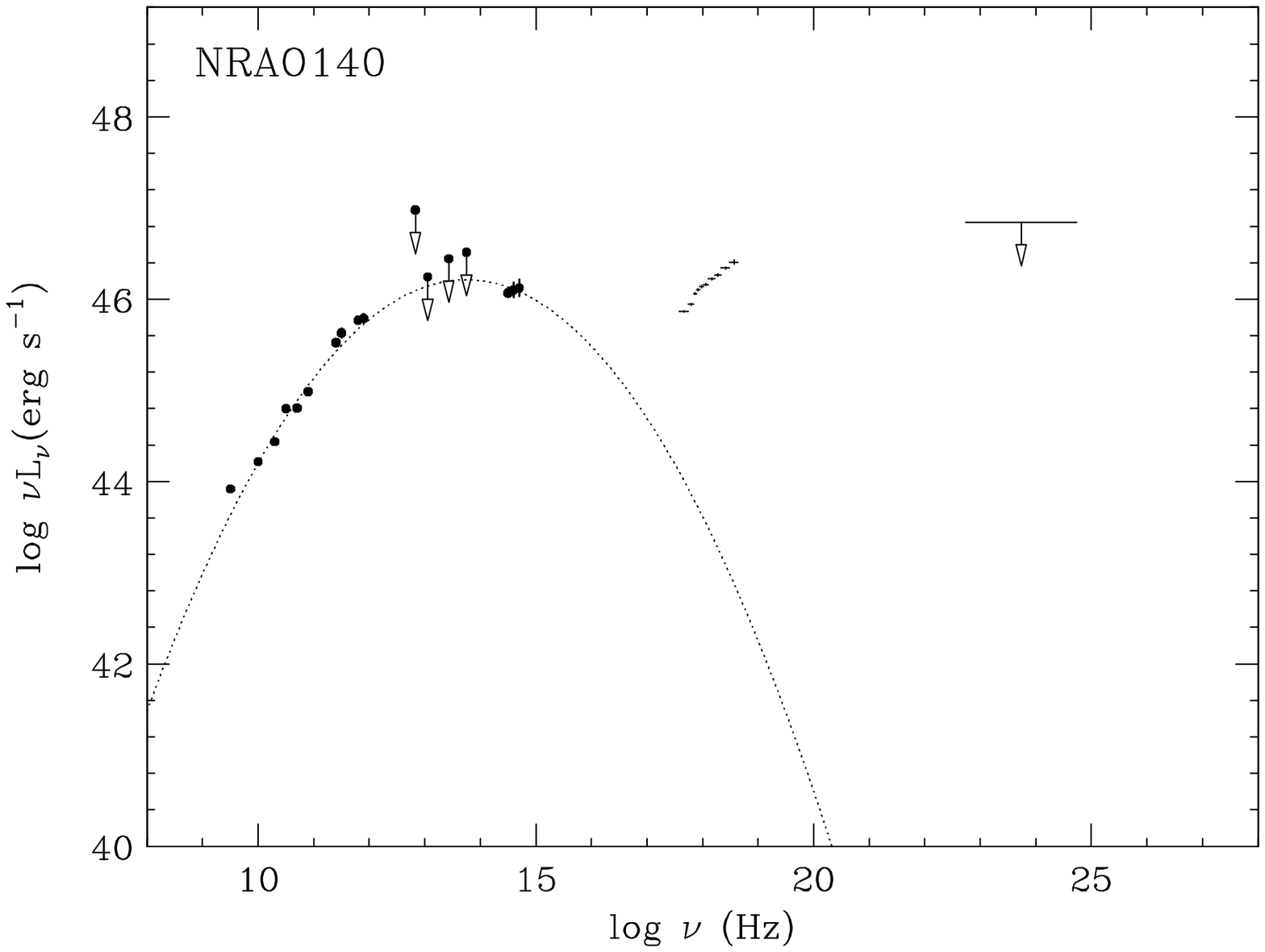}
\plotone{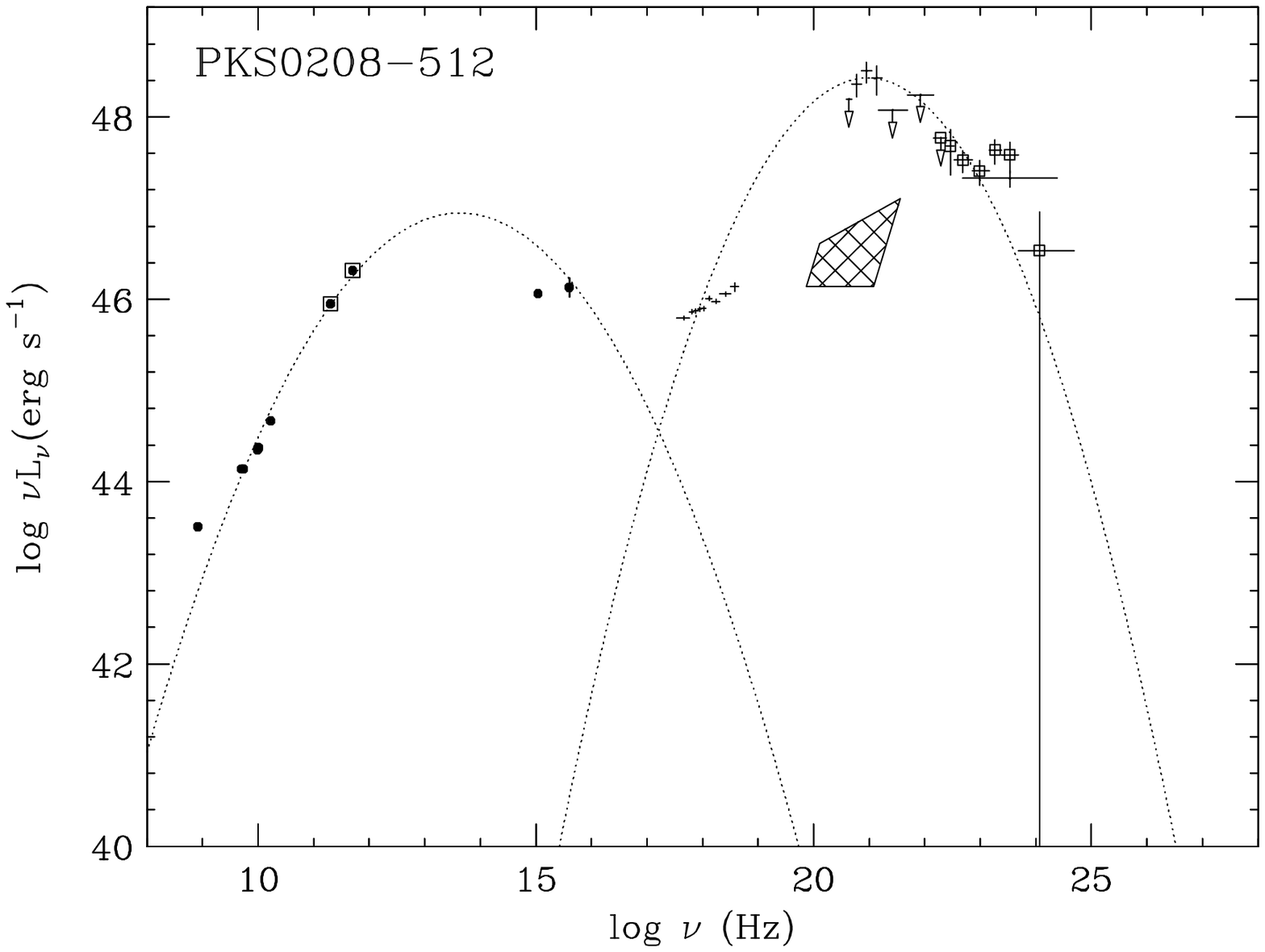}

\plotone{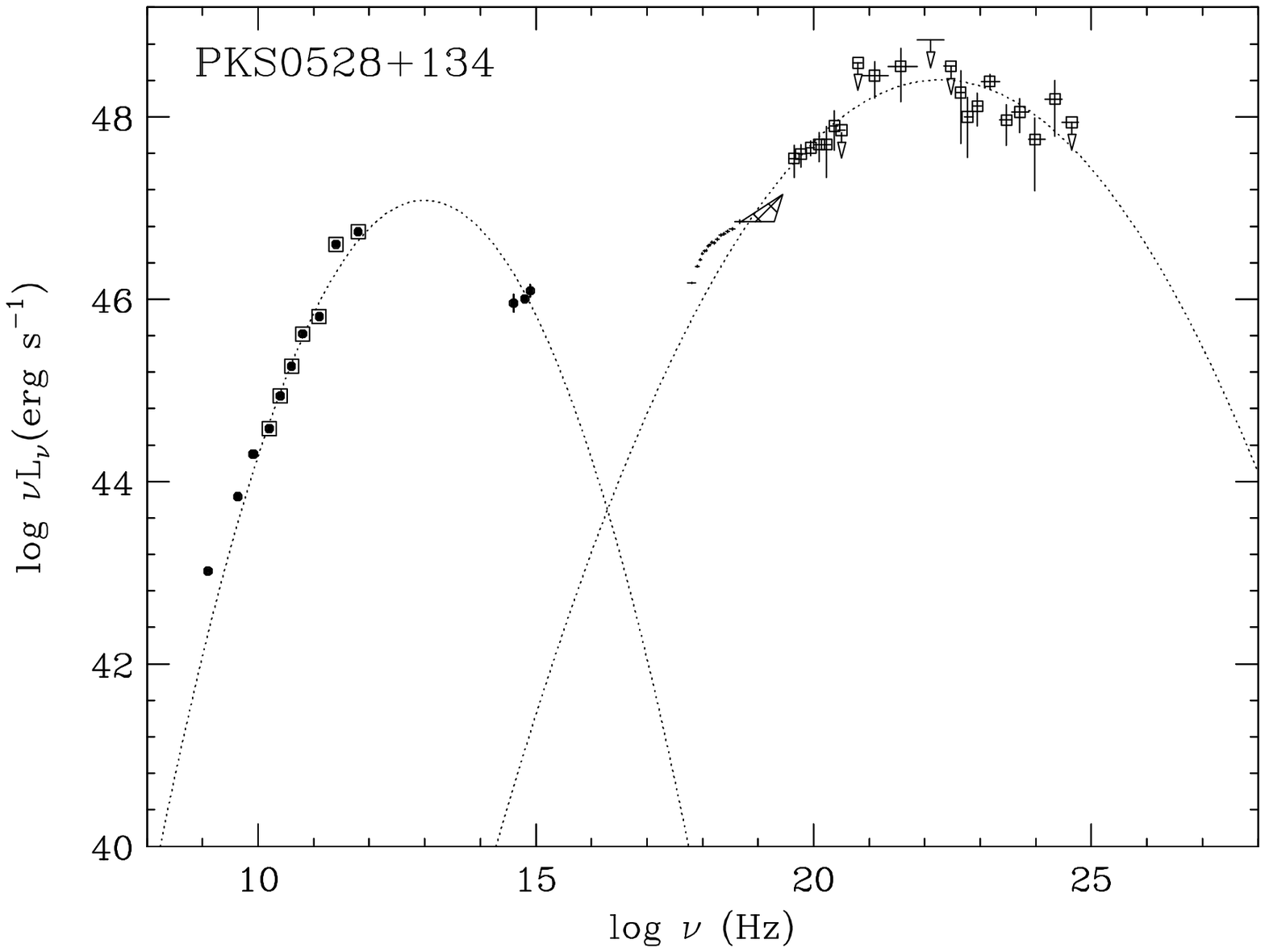}
\figcaption{
Multiband spectra of blazars, showing the 
distribution of flux versus the rest-frame frequency.
Square symbols show simultaneous observations with {\sl ASCA}, although 
for radio data they include quasi-simultaneous observations.
The dashed line is the polynomial fit to the spectrum.
The X--ray spectra are corrected for absorption using the Galactic column 
density. Hatched region in QHBs corresponds to the estimated range 
of $\nu_{SSC}$ and $L_{SSC}$.  For 3C~279, we plot the {\sl ASCA} 
as well as the \ginga X--ray spectra;  for this object, 
the squares represent observations contemporaneous 
with the plotted \ginga data (see also \protect{\cite{maraschi94b}}).  
The data from the radio to optical bands were mainly taken from NED database.
The radio data were also taken from UMRAO database, 
and Ter\"asranta \& Tornikoski in private communication.
We also used data from \protect{\cite{bersanelli92}} (optical), 
\protect{\cite{pian93}} (UV), 
\protect{\cite{falomo94}} (optical), \protect{\cite{giommi95}} (radio to X--ray),
\protect{\cite{impey88}} (IR),
\protect{\cite{neugebauer86}} (IR), 
\protect{\cite{mcnaron95}} (OSSE, COMPTEL, EGRET),
\protect{\cite{bloom94}} (radio, mm, IR), 
\protect{\cite{fichtel94}} (EGRET upper limit).
Additional data for each object were taken from 
H0323+022 (\protect{\cite{feigelson86}}; \protect{\cite{jannuzi93}}; 
\protect{\cite{falomo93}}), 
H0414+009 (\protect{\cite{mchardy92}}),  
PKS~2155--304 (\protect{\cite{vestrand95}}),  
Mkn~421 (\protect{\cite{macomb95}}),
Mkn~501 (\protect{\cite{kataoka98}}),
H1426+428 (\protect{\cite{remillard89}}), 
OJ287 (\protect{\cite{landau83}}; \protect{\cite{idesawa97}}; \protect{\cite{shrader96}}), 
AO0235+164 (Madejski et al. 1996), 
PKS~0735+178 (\protect{\cite{nolan96}}), 
0716+714 (\protect{\cite{lin95}}), 
3C~273 (\protect{\cite{montigny93}}; \protect{\cite{montigny97}}; 
\rosat: MPE Annual Report 1993), 
3C~279 (\protect{\cite{hartman96}}), 
PKS~0208--512 (\protect{\cite{veron93}}; \protect{\cite{blom95}}; \protect{\cite{stacy96}}),
PKS~0528+134 (\protect{\cite{rieke82}}; \protect{\cite{collmar98}}), 
4C~38.41 (\protect{\cite{landau86}}; \protect{\cite{mattox93}})\label{fig2}}

\clearpage

\epsscale{1.0}
\plotone{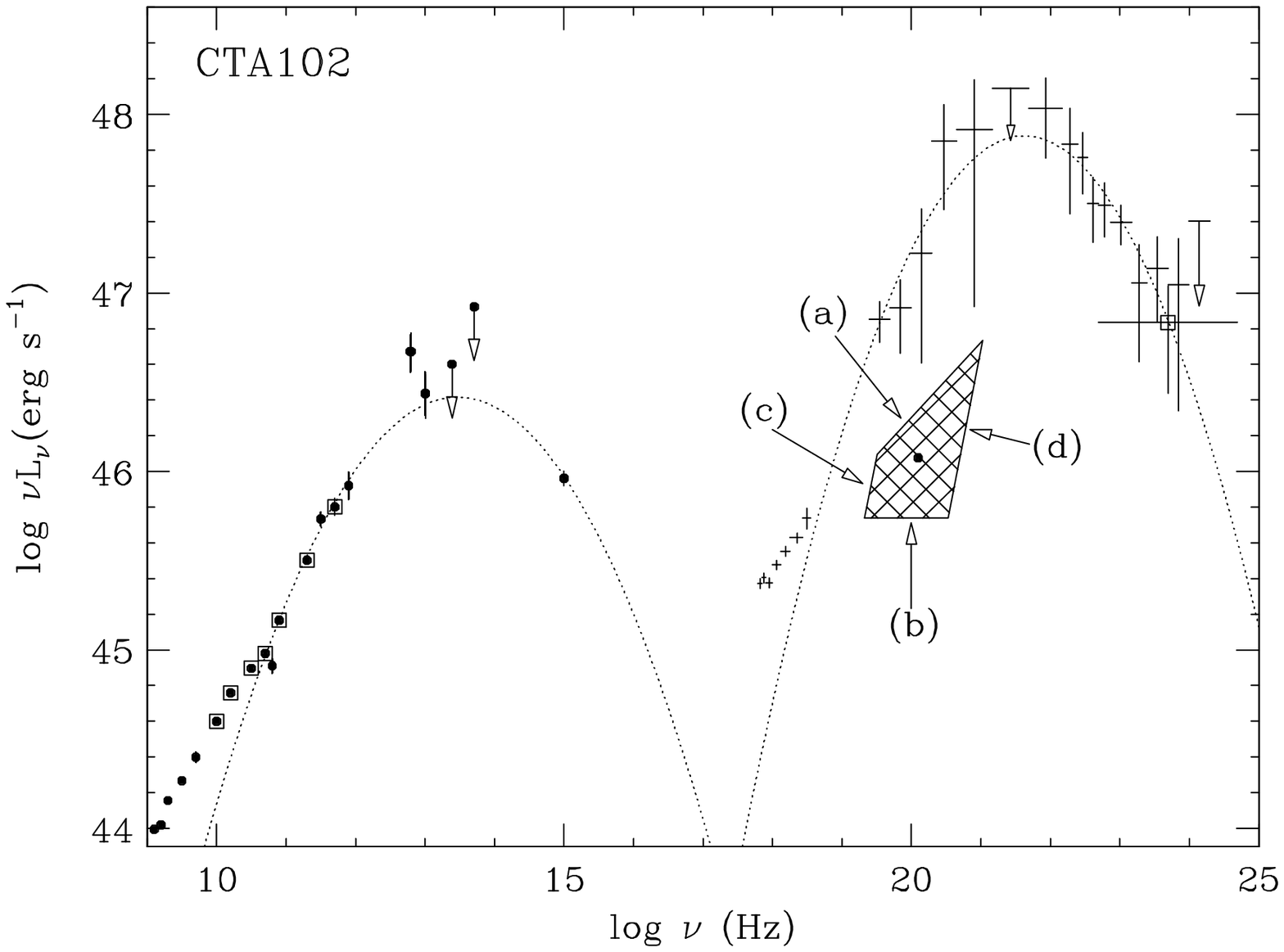}
\figcaption{
Multiband spectrum of CTA~102, where we illustrate in more detail the 
procedure for choosing input parameters for the model used in the text.  
Hatched region corresponds to the estimated range of $\nu_{SSC}$ and $L_{SSC}$.
The lines bounding the shaded area are:  
(a) a linear extrapolation of the 
\asca spectrum, (b) the highest value of $\nu F(\nu )$ measured by {\sl ASCA}, 
(c) ($\nu_{SSC}$, $L_{SSC}$) for $\delta$=5, and 
(d) ($\nu_{SSC}$, $L_{SSC}$) for $\delta$=20.
The filled symbol in the hatched region 
shows the point that is used in the calculation of 
magnetic field $B$ and electron Lorentz factor $\gamma_b$.  
The data used are from sources as given in Fig. 2;  
in addition, we used \protect{\cite{nolan93}} 
and \protect{\cite{padovani92}}.\label{fig3}}

\clearpage

\epsscale{0.7}
\plotone{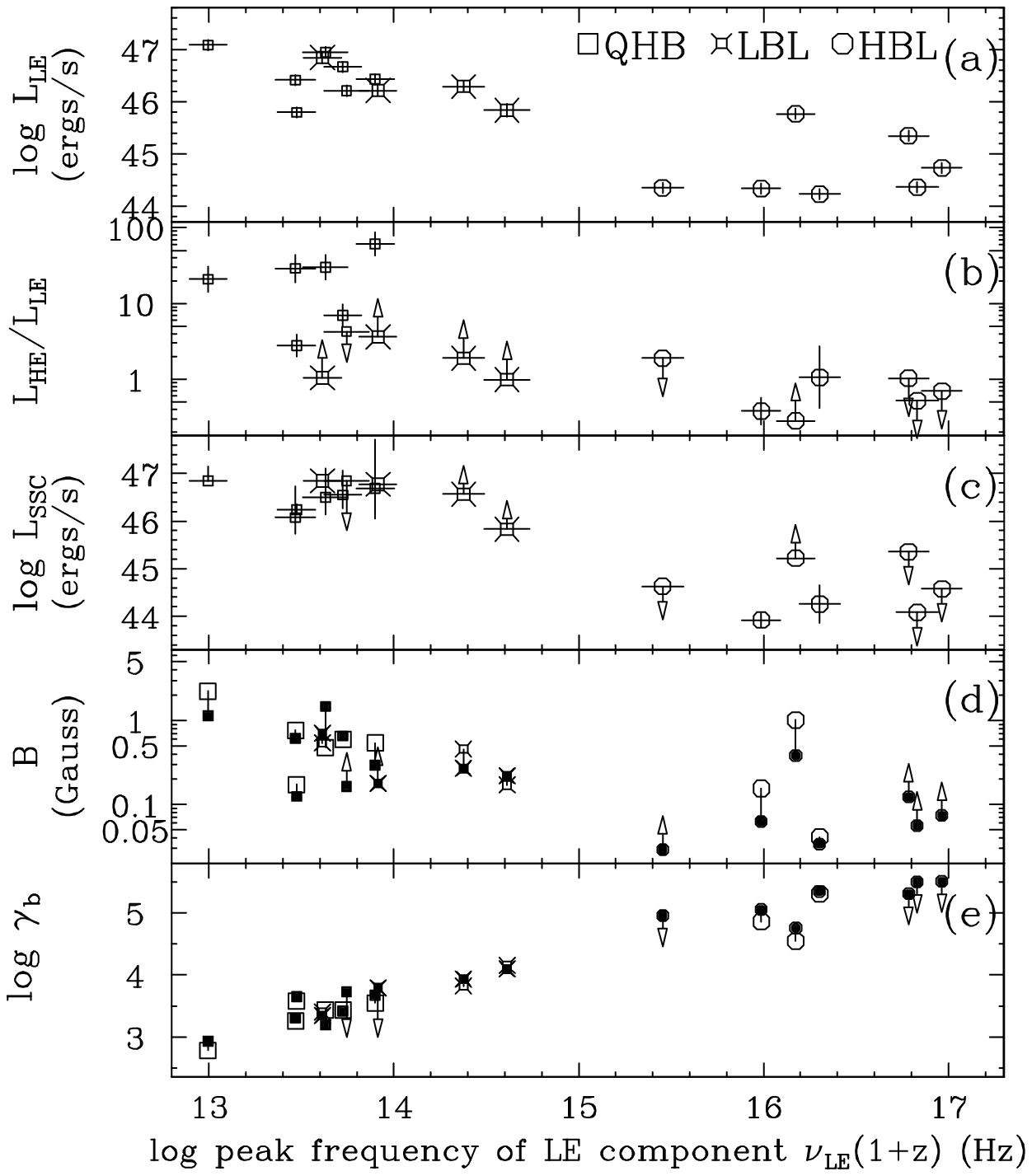}
\figcaption{
The distribution of parameters derived from the multi-band analysis 
as a function of the rest-frame peak frequency of the LE component.
(a) LE luminosity in the observer frame ($L_{LE}$);  
(b) the ratio $L_{HE}$/$L_{LE}$ 
(c) SSC luminosity in the observer frame ($L_{SSC}$);  
(d) magnetic field ($B$); and
(e) Lorentz factor of electrons radiating at the peak 
of the $\nu F(\nu)$ spectrum ($\gamma_{b}$).
In the calculation, we use the beaming factor 
$\delta$ = 10.  The size $R$ is estimated from the observed time variability 
from Table 2 (open symbol).  We also plot values calculated with 
$R$ = 0.01 pc (filled symbol). 
For HBLs, QHB 3C~273, and two LBLs PKS~0735+178, 0716+714, 
we use the $L_{HE}$ as $L_{SSC}$.
For LBL AO0235+164 we assume $L_{SSC}$ = $L_{LE}$. 
For QHBs except 3C~273, $L_{SSC}$ is estimated by extrapolating 
the \asca spectrum.  
Downward arrows in (b),(c),(e) and upward ones in (d) show 
sources that are not detected in the $\gamma$--ray band. 
Upward arrows in (b),(c),(e) and downward ones in (d) show 
sources whose $\gamma$--ray emission is detected, 
but the peak of the HE component is not determined.\label{fig4}}

\end{document}